\documentclass[preprint2]{aastex}

\usepackage{natbib}
\usepackage{graphicx, subfigure}
\usepackage{multirow}
\usepackage{color}
\usepackage{amsmath} % or simply amstext

\shorttitle{ A companion at the brown dwarf/planet limit to a Tucana-Horologium M dwarf}
\shortauthors{Artigau et al.}

\begin{document}

\title{ BANYAN. VI. Discovery of a companion at the brown dwarf/planet-mass limit to a Tucana-Horologium M dwarf}

\author{ {\'E}tienne Artigau\altaffilmark{1}, Jonathan Gagn\'e \altaffilmark{1}, Jacqueline Faherty\altaffilmark{2}, Lison Malo\altaffilmark{3} , Marie-Eve Naud\altaffilmark{1}, Ren\'e Doyon\altaffilmark{1}, David Lafreni{\`e}re\altaffilmark{1}, Yuri Beletsky\altaffilmark{4} }

\altaffiltext{1}{Institut de Recherche sur les Exoplan{\`e}tes (IREx), D\'epartement de Physique, Universit\'e de Montr\'eal, C.P. 6128, Succ. Centre-Ville, Montr\'eal, QC, H3C 3J7, Canada}
\altaffiltext{2}{Department of Terrestrial Magnetism, Carnegie Institution of Washington, Washington, DC 20015, USA}
\altaffiltext{3}{Canada-France-Hawaii Telescope Corporation, 65-1238 Mamalahoa Highway, Kamuela, HI 96743, USA}
\altaffiltext{4}{Las Campanas Observatory, Carnegie Institution of Washington, Colina el Pino, Casilla 601, La Serena, Chile}

\email{Send correspondence to artigau@astro.umontreal.ca}

\begin{abstract}
We report the discovery of a substellar companion to 2MASS J02192210--3925225, a young M6\,$\gamma$ candidate member of the Tucana-Horologium association  ($30-40$~Myr). This  L4\,$\gamma$ companion has been discovered with seeing-limited direct imaging observations; at a $4^{\prime\prime}$ separation ($160$\,AU) and a modest contrast ratio, it joins the very short list of young low-mass companions amenable to study without the aid of adaptive optics, enabling its characterization with a much wider suite of instruments than is possible for companions uncovered by high-contrast imaging surveys. With a model-dependent mass of 12--15\,M$_{\rm Jup}$, it straddles the boundary between the planet and brown dwarf mass regimes. We present near-infrared spectroscopy of this companion and compare it to various similar objects uncovered in the last few years.  The J0219--3925 system falls in a sparsely populated part of the host mass versus mass ratio diagram for binaries; the dearth of known similar companions may be due to observational biases in previous low-mass companion searches.
\end{abstract}

\keywords{}

\section{Introduction}
The spectacular discoveries brought on by high-contrast imaging of exoplanets around nearby stars (e.g., HR8799's system, \citealt{Marois:2008, Marois:2010}$;\beta$ Pictoris b, \citealt{Lagrange2010}; Fomalhaut's companion, \citealt{Kalas2008} ) eclipse a fact that is often overlooked: most of the region around a star where a planet would be gravitationally bound for Gyrs is readily accessible to seeing-limited observations for nearby stars. A handful of planetary-mass objects have indeed been uncovered in wide-field surveys, such as GU Psc b and Ross 458(AB)c, respectively 2000 and 1200\,AU from their hosts \citep{Goldman:2010, Burningham2011, Naud:2014}, and the presence of companions on such wide orbits presents a significant challenge to planetary formation models.

 These companions are on orbits too large for in situ formation by either core accretion \citep{Pollack1996, Alibert2005} or gravitational instability within protoplanetary disks \citep{Cameron1978, Boss1997}. \citet{Vorobyov:2013} established that wide companions may form in wide orbits within gravitationally unstable protoplanetary disks, but only around massive ($>0.7$\,M$_\odot$) hosts with massive ($>0.2$\,M$_\odot$) protoplanetary disks, and cannot account for the discovery of such companions around K stars or later-type hosts. Turbulent fragmentation of a pre-stellar core \citep{Padoan:2002} is a viable alternative to explain the existence of such systems. Outward migration through planet-planet interaction is also plausible, but remains to be tested as a credible mechanism at such separations and mass ratios \citep{Veras:2009}. Statistical analyses of the mass, age and separation of planets uncovered through high-contrast imaging campaigns suggest that these companions constitute a low-mass tail of the brown dwarf distribution and are therefore the results of disk or cloud fragmentation. The dearth of low-mass ($<5$\,M$_J$) companions found by direct imaging surveys with adaptive optics (AO) can be explained in that context \citep{Brandt:2014a}. This analysis mostly contrains the occurence of companions out to 100-200\,AU; whether this paucity of companions holds for more distant brown dwarfs and planetary-mass companions remains to be determined in a statistical framework. Overall, much work needs to be done to understand the origin and demographics of these distant companions, both theoretically and observationally.

These distant companions have distinct advantages compared to either isolated planetary-mass objects or planets detected through AO surveys. The presence of a host star allows a cross-calibration of the properties of the planet (parallax, mass, age, metallicity, membership to a young group), and the projected separation to that host is sufficient to allow direct study through means not compatible with extreme-AO, such as accurate spectro-photometry, high-resolution spectroscopy, optical imaging, etc. Indeed, the detailed analysis performed with intermediate-resolution spectro-photometry on GU Psc b and Ross 458(AB)c will be impossible to obtain in the near future for exoplanets uncovered by AO surveys.

As distant companions of nearby young stars provide important benchmarks to understand self-luminous gas giants, we undertook various seeing-limited observations using the SIMON near-infrared spectro-imager \citep{Doyon:2000b} at the CTIO-1.5\,m telescope and GMOS-S at Gemini South in order to identify such new objects through their distinctive far-red and near-infrared colors. We report here a first discovery from this survey; a co-moving companion to 2MASS J02192210--3925225 (J0219--3925), an M6\,$\gamma$ candidate member of the Tucana-Horologium association (THA) that has a low-gravity L4\,$\gamma$  companion.

The survey that led to this discovery is described in section~\ref{discovery}. The discovery and photometric follow-ups of the companion and its host are described in section~\ref{observations}. Results are described in section~\ref{results} and discussed in section~\ref{discussion}.

\section{The survey}

\label{discovery}

J0219--3925 has been observed as part of a survey of 300 stars conducted at the CTIO-1.5\,m telescope with the SIMON. This survey was undertaken following the discovery of the planetary-mass companion around the M3 AB Doradus member GU Psc \citep{Naud:2014}, in an attempt to identify additional comparable planetary-mass companions and assess their overall frequency. The target sample consists of both confirmed and strong members of young moving groups  within 70\,pc ($<120$\,Myr; $\beta$ Pictoris, AB Doradus, THA, Columba, Carina and Argus), and it was assembled from objects in  \citet{Malo:2013} and \citet{Gagne:2015}. For each star, the Bayesian Analysis for Nearby Young AssociatioNs II tool (BANYAN\,II; \citealt{Gagne:2014d}) provided a statistical distance estimate, allowing us to derive a projected distance for putative companions. 

By combining SIMON $J$-band imaging with WISE photometry \citep{Wright:2010}, we identified objects that had a similar position to brown dwarfs or known very-low gravity L dwarfs in an M$_{W1}$ vs $J-W2$ diagram, but that did not have optical digitized sky survey (DSS) counterparts. 

In a way similar to the survey that allowed the discovery of GU Psc\,b, candidates were then followed-up with Gemini-South using deep $i$ and $z$ band imaging to confirm the very red $i-z$ color expected for an L or T dwarf. The selection methods will be detailed in a future paper (Artigau et al. in preparation), as refinement of selection methods and follow-up of candidates are still ongoing. 

As the SIMON observations have significantly better resolution than 2MASS ($\sim$$1\arcsec$ versus $\sim$$2\arcsec$ full-width at half maximum) and are significantly deeper ($10$\,$\sigma$ at $J\sim18$), they provide an opportunity to identify relatively tight ($2-6\arcsec$) overlooked companions. We performed a radial profile subtraction on all primaries and visually inspected the residuals. Through this process, a single candidate was identified, at an angular distance of about $4\arcsec$ from the M6 star J0219--3925. No other star within our sample presented a similar candidate companions within in a $2-6\arcsec$ annulus down to a contrast of $\Delta J<5$.

\section{Observation and reduction}
\label{observations}

\begin{table*}[!htbp]
\begin{center}
 \caption{Imaging and spectroscopy datasets for the J0219--3925 system}
\begin{tabular}{ l ccccc }
\hline
\hline
Instrument & Date & Filter & Number &Per-frame &Comment \\
 & & & of exposures&exposure time & where applicable \\
 & & &                     & &Gemini program ID \\

\hline
GMOS-S & 17/09/2012& $g$ & 1	&10.5\,s & Spectroscopy acquisition\\
		&		&		&	&		&GS-2012B-Q-70\\
GMOS-S & 02/11/2012& $i$ & 1	&10.5\,s & Spectroscopy acquisition\\
		&		&		&	&		&GS-2012B-Q-70\\
SIMON & 05/11/2013& $J$ & 60 & 30\,s & Discovery imaging\\
SIMON & 12/02/2014& $H$ & 20 & 30\,s & Photometric follow-up\\
SIMON 	& 12/02/2014& $K_s$ & 20 & 30\,s & Photometric follow-up\\

F2		& 28/10/2013& $J$ & 1	&5\,s& Spectroscopy acquisition\\
		&		&		&	&		&GS-2013B-Q-79\\

FIRE		& 13/02/2014 & $\cdots$& 2 & 306\,s & J0219--3925, spectroscopy \\
FIRE		& 13/02/2014 & $\cdots$& 4 &909\,s & J0219--3925\,B, spectroscopy\\

F2		& 04/09/2014 & $J$ & 27 & 54\,s & Imaging for astrometry\\
		&		&		&	&		&GS-2014B-Q-72\\

\hline
\hline
\end{tabular}
\end{center}
\end{table*}

\subsection{Imaging}

\label{simon}
The $J$-band discovery dataset of J0219--3925 was obtained with SIMON on 2013 November 5 at the CTIO, and follow-up $H$ and $K_s$-band imaging were obtained on 2014 February 2, with the same instrument and telescope. For all 3 imaging sequences, we employed a 4-point dither pattern along the corners of a $2\arcmin\times2\arcmin$ square. At each dither position, 15 images ($J$) and 5 images ($H$ and $K_s$) were taken. All sequences used a 30\,s per-frame exposure time. The images were sky-subtracted using a sky frame constructed from the median combination of all science images taken on that night within that band. Flat-fielding was performed with a flat constructed from images of a flat screen. The astrometric solution was performed using a cross-match of the 2MASS catalog with field stars. All images were registered and median-combined to produce the final science frame.

The contrast ratio between the two components was determined by performing a PSF fitting of the two J0219--3925 components using two isolated bright and nearby field stars ($<2\farcm5$, $J>13.5$) as input PSFs. The magnitude of J0219--3925\,B was then determined from the 2MASS magnitude of its host and its contrast ratio within each near-infrared bandpass. The presence of the companion is unlikely to have significantly affected the 2MASS magnitudes of its host, as it contributes only 2 to 4\% of the total integrated flux and is resolved ($\sim2$ FWHM).

\begin{figure}[!htbp]
\includegraphics[width=0.42\textwidth]{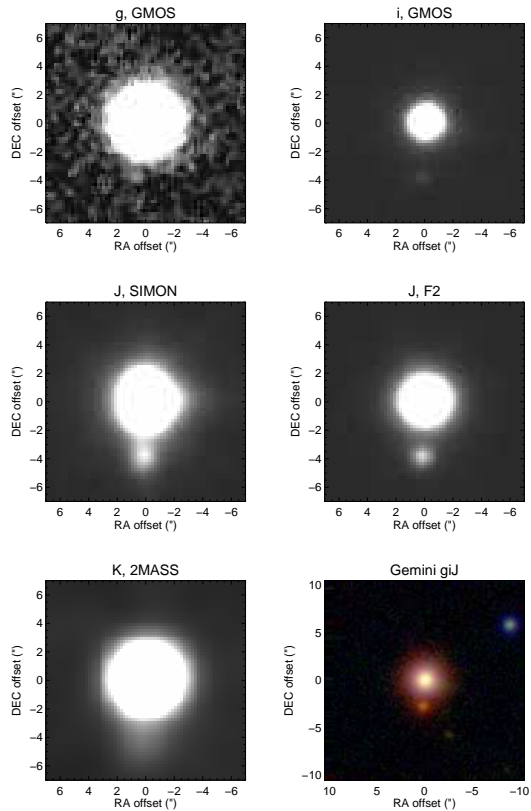}

\caption{J0219--3925 system in SIMON, Flamingos\,2 (F2), GMOS and 2MASS imagery. Despite modest constraints provided by the 2MASS image on the position of J0219--3925\,B, the $\sim$15\,yr delay between these archival observations and our discovery images provides significant constraints on the common proper motion (see section~\ref{commonpm}). For all images, East is left, North is up. The color image is a combination of $g$, $i$ and F2 $J$-band images respectively coded as blue, green and red with arbitrary scaling. One readily sees that J0219--3925\,B is much redder than its host. \label{imaging}}
\end{figure}

Upon discovery of a possible faint companion to J0219--3925, we noticed that images of that field usable for science were present in the Gemini science archive (See Figure~\ref{imaging}). GMOS-South ($i$ and $g$-band) and Flamingos-II ($J$-band) spectroscopic acquisition images of the host star have been obtained as follow-up observations of the BANYAN All-Sky Survey (BASS;  \citealt{Gagne:2015}). The $\Delta i=6.22$ contrast and non-detection in $g$ ($\Delta g>5.5$) pointed toward an object much redder than the mid-M host, prompting for a dedicated spectroscopic follow-up of the companion. Inspection of the $K$-band 2MASS images taken in 1999 shows that the companion is marginally detected and provided valuable constraints on the common proper motion of the pair (see Section~\ref{commonpm} and Table\,\ref{observations}).

\subsection{Spectroscopy}

\label{fire}
On 2014 February 13, we obtained near-infrared spectroscopic observations with the Folded-Port Infrared Echelette (FIRE; \citealt{Simcoe:2013}) at the Magellan 6.5-m telescope. FIRE was used in its high-resolution echellette mode, providing a spectral resolution $R\sim5000$ continuously over the entire near-infrared domain ($0.8-2.4\mu$m). We used a slit width of $0\farcs6$ under a $1\farcs2$ seeing and an airmass of $1.3-1.6$ (J0219--3925\,B) and $1.8-1.9$ (J0219--3925\,A). The reference star for both objects was observed at the same airmasses. For J0219--3925\,B, we used an ABBA dither pattern for improved sky subtraction on this relatively faint target, while for the brighter J0219--3925\,A and reference star, the two exposures were taken with an AB dither pattern.

\begin{figure*}[!tbp]
\includegraphics[width=0.49\textwidth]{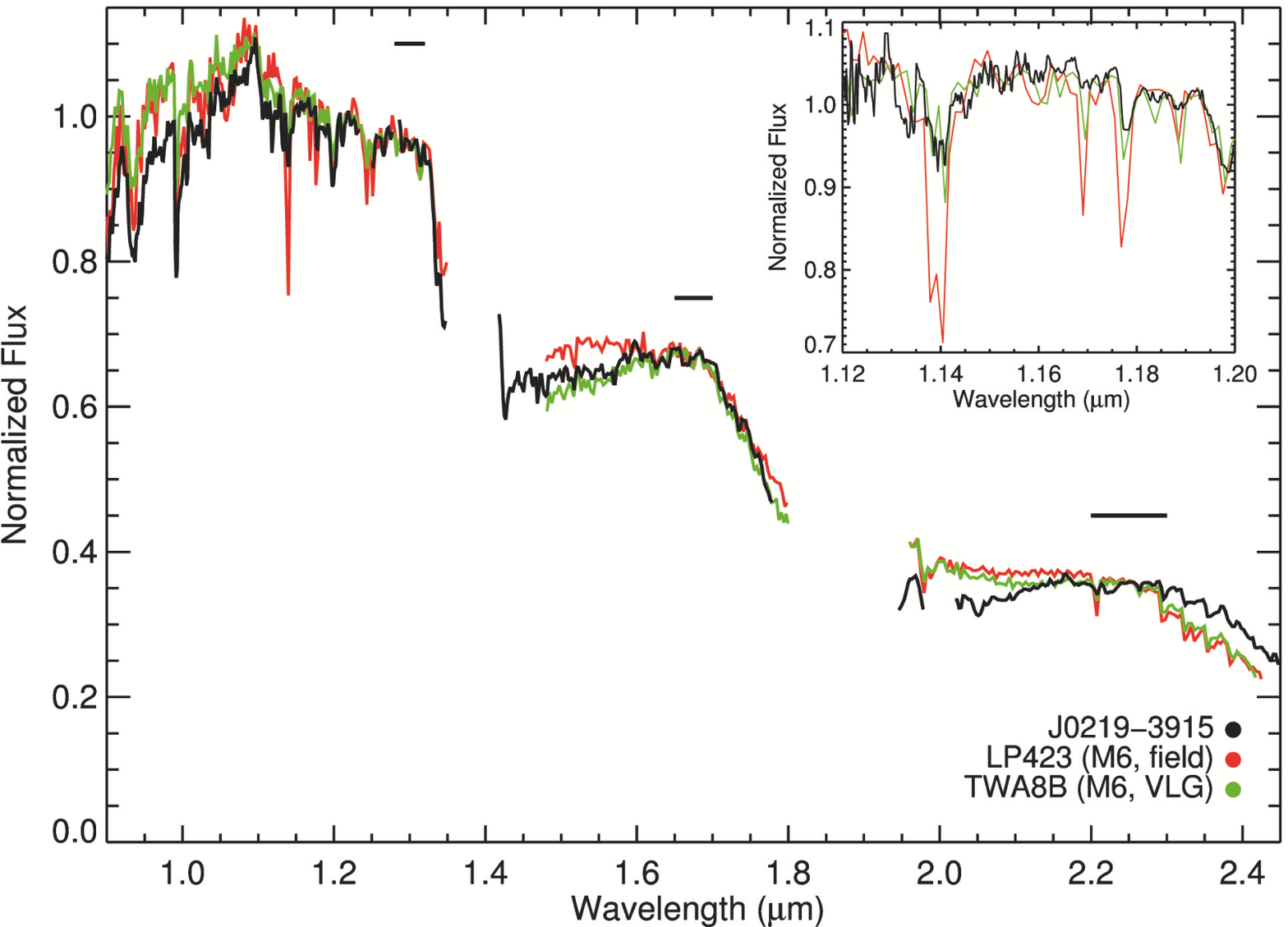}
\includegraphics[width=0.49\textwidth]{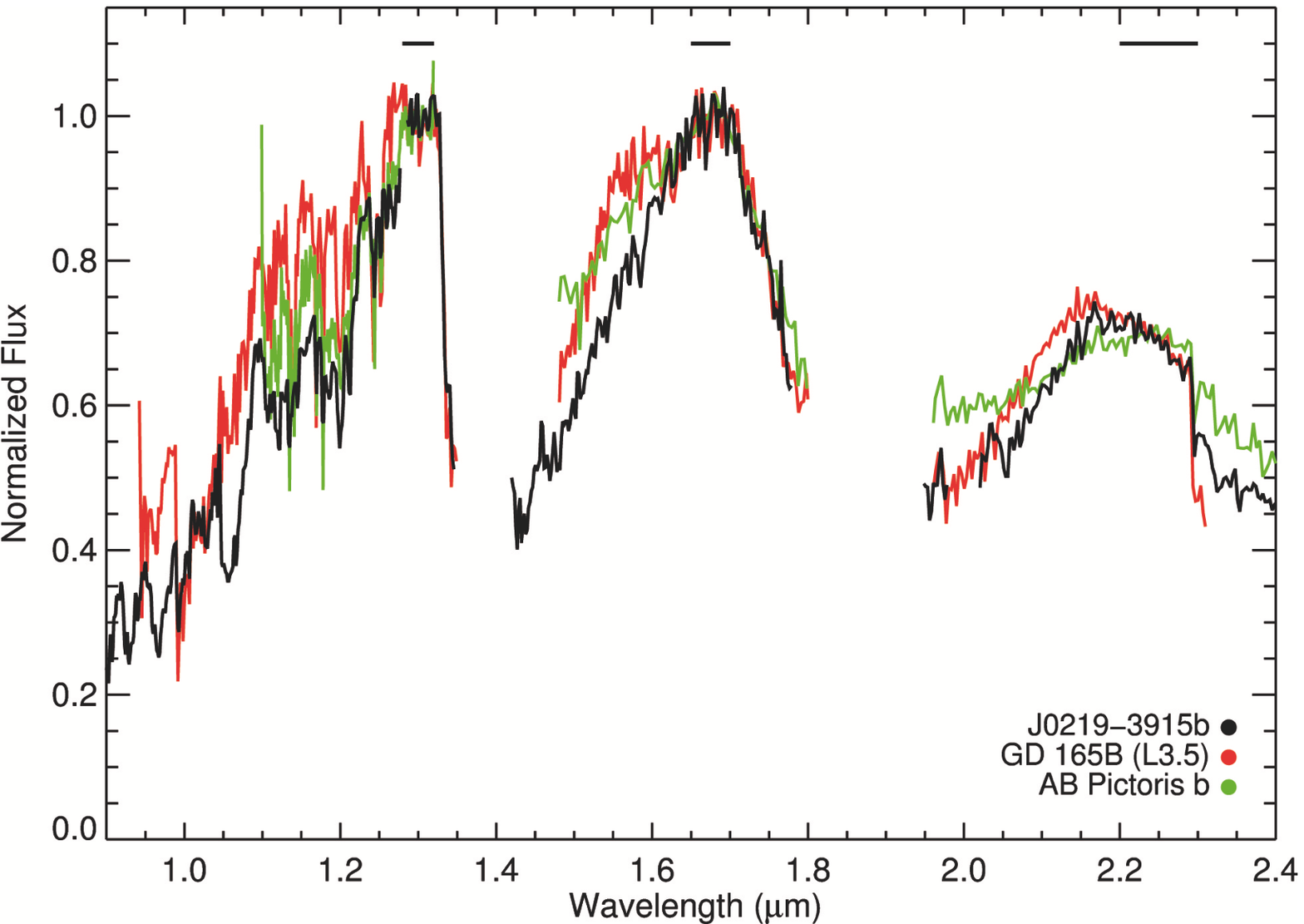}

\caption{ $(left)$ Spectra of J0219--3925, compared to a field M6 dwarf (LP423) and an M6\,$\gamma$ (TWA8B). Both J0219--3925 and TWA8B show an $H$-band spectrum that is more peaked that that of the higher gravity LP423. The inset shows that full resolution spectra in the $1.12-1.20\mu$m domain; the higher gravity LP423 shows much stronger Na\textsc{I} absorption at $1.138\mu$m and a stronger K\textsc{I} doublet at $\sim$ $1.17\mu$m. $(right)$ Spectra of J0219--3925\,B, the young companion AB Pictoris\,b (data from \citealt{Bonnefoy:2014}) and the field L3.5 dwarf GD 165 B (data from \citealt{McLean:2003}). The $H$-band flux of J0219--3925 is noticeably more peaked than the two others, pointing toward a very low gravity (compare with middle panels in Figure~\ref{spfit}). $(both)$ Flux within each photometric bandpass has been normalized to that of J0219--3925\,A or J0219--3925\,B to highlight differences in the shape of the SED within each bandpass; normalization intervals are shown as straight black lines above spectra. \label{sp1} }
\end{figure*}

 Data reduction was performed using the standard FIREHOSE pipeline. Flat-field correction was performed using flat frames derived from dome flat images and telluric absorption at the time observation was derived from an A0V star (HD 17683) observation taken immediately before (J0219--3925\,B) and after (J0219--3925\,A) the science integration. The spectrum of the host star and its companion are compared to field and low-gravity objects of comparable spectral types in Figure~\ref{sp1}. The spectrum of J0219--3925\,B has an average S/N of $\sim30$ when sampled at an $R=800$ resolution.

\section{Results}
\label{results}

\begin{table*}[!htbp]
\caption{{Properties of the J0219Ñ3925 system}}
\begin{tabular}{ l cc }

\hline
\hline

Short name & J0219--3925& J0219--3925\,B\\
2MASS name & \multicolumn{2}{c}{02192210--3925225 } \\

$\mu_{\alpha}$$^a$ (mas/yr)&$107.37\pm2.27$ & \\
$\mu_{\delta}$$^a$ (mas/yr)&$-34.95\pm1.65$ & \\

Statistical distance$^b$ &\multicolumn{2}{c}{$39.4\pm2.6$\,pc } \\
Position angle & \multicolumn{2}{c}{ $173.9\pm0.2^\circ$ }\\
Separation & \multicolumn{2}{c}{$ 3.96\pm0.02$\arcsec}\\

Sky-plane separation &\multicolumn{2}{c}{ $156\pm10$\,AU}\\
$B$&18.50 &$\cdots$ \\
$\Delta g^c$ & \multicolumn{2}{c}{$>5.5$}\\
$\Delta i^c$ & \multicolumn{2}{c}{$6.6$}\\
$R$&15.23 & $\cdots$\\
$I$$^d$&$13.42\pm0.03$ &$\cdots$ \\ 
$J_{2Mass}$$^e$ & $11.381\pm0.026$ & $\cdots$ \\
$J_{MKO}$$^f$ & $11.321\pm0.026$ & $15.54\pm0.10$\\
$H_{2Mass}$$^e$ & $10.811\pm0.027$ & $\cdots$ \\
$H_{MKO}$$^f$ & $10.855\pm0.027$ & $14.63\pm0.10$\\
$K_{2Mass}$$^e$ & $10.404\pm0.025$ &$\cdots$ \\
$K_{MKO}$$^f$ & $10.444\pm0.025$ &$13.82\pm0.10$\\
$W1$$^g$ &\multicolumn{2}{c}{ $10.148\pm0.023$} \\
$W2$$^g$ &\multicolumn{2}{c}{ $9.901\pm0.020$ } \\
$W3$$^g$ &\multicolumn{2}{c}{ $9.614\pm0.037$ } \\
$v_{rad}$ & $10.6\pm0.7^{\rm g}$ &$\cdots$ \\
$v \sin i$ & $6.5\pm0.4^{\rm g}$ & $\cdots$\\
SpT & M6\,$\gamma^{\rm h}$ & L4\,$\gamma$\\
EW H$\alpha$ (\AA) & -7.02 & $\cdots$\\
EW Li$_{6708}$ (\AA) & 639.9 & $\cdots$\\
FeH$_z$ & $1.068\pm0.001$ & $1.040\pm0.002$\\
FeH$_J$ & $1.07\pm0.01$ & $1.20\pm0.02$\\
K\textsc{i} 1.169$\mu$m & $0.17\pm0.08$ & $0.6\pm0.4$\\
K\textsc{i} 1.177$\mu$m& $1.25\pm0.06$  & $0.71\pm0.32$\\
K\textsc{i} 1.244$\mu$m & $\cdots^i$ & $3.0\pm0.5$\\
Na\textsc{i} 1.138$\mu$m & $2.58\pm0.08$ & $2.4\pm0.5$\\
$H$-Cont & $0.995\pm0.001$ & $0.963\pm0.001$\\
VO$_z$ &$1.001\pm0.001$ & $1.372\pm0.004$\\

\hline\hline\\
\end{tabular}

\tablenotetext{a}{\citealt{Girard2011}}
\tablenotetext{b}{See \citealt{Gagne:2014d} for a discussion on the definition and the accuracy of the statistical distance.}
\tablenotetext{c}{Only contrasts were derived in $g$ and $i$, and not magnitudes measurements as the images were not taken under photometric conditions.}
\tablenotetext{d}{\citealt{DENIS:2005}}
\tablenotetext{e}{\citealt{Skrutskie:2006}}
\tablenotetext{f}{The  2MASS to MKO  transform has been determined for J0219--3925 from its FIRE spectrum. The companion's contrast ratio has been determined with SIMON that uses MKO filters.}
\tablenotetext{g}{\citealt{Cutri:2013}}
\tablenotetext{h}{\citealt{Kraus:2014} give a M5.9 spectral type from SED fitting and a spectroscopic spectral type of M4.9.}
\tablenotetext{i}{The spectrum of J0219--3925\,B around the KI 1.244$\mu$m is strongly affected by bad pixels on the science array.}

\end{table*}

\subsection{Host star properties and membership}

\label{host}

J0219--3925 has been identified as a member of the THA by two teams independently. First, it is part of the 129 new late-type (K3 and later) THA members identified by \citealt{Kraus:2014}; their membership being established by Li and RV measurements, but lack parallaxes and is based on spectroscopic distances. Spectroscopic fitting of optical spectrum and overall SED of J0219--3925 respectively lead to a spectral type estimate of M$4.9\pm1.0$ and M$5.9\pm0.3$; these values are consistent with the near-infrared spectral type of M6\,$\gamma$ determined here (see section~\ref{gravity}). The radial velocity they measure ($10.6\pm0.7$\,km\,s$^{-1}$) agrees within $0.71$\,km\,s$^{-1}$ with the expected radial velocity of a THA member in that line of sight. This is below the internal dispersion of THA velocities ($\sim$$1$\,km\,s$^{-1}$) and well below the cutoff in velocity difference between members and non-members in their analysis ($\pm3$\,km\,s$^{-1}$).

The host star J0219--3925 was identified independently as a candidate member of THA as part of the BASS survey. \emph{2MASS}  and  \emph{AllWISE} photometry and astrometry were used to derive its membership probability  to several moving groups in the Solar neighborhood. The BANYAN\,II tool compare the input parameters to spatial, kinematic and photometric models of young moving groups and the field population using a naive Bayesian classifier. We refer the reader to \cite{Gagne:2014d} for details on this analysis. As J0219--3925 had been identified as a high probability candidate of THA, it was included in our search for distant companions before the publication of the bulk of the survey. The BANYAN\,II tool gives a P=99.34\% probability that J0219--3925 is a member of THA. When such measurements are available, the BANYAN~II tool can use the radial velocity and/or parallax to constrain membership probability further. Using the radial velocity of $10.6\pm0.7$\,km\,s$^{-1}$ measured by \citet{Kraus:2014}, the THA membership probability of the host star increases to P=99.94\%. 

We can set additional constraints on the age of this object  since it displays lithium \citep{Kraus:2014} and typical signs of low-gravity in its NIR spectrum (see left panel of figure~\ref{sp1} and Section~\ref{gravity}). Within that spectral type, the presence of lithium (EW$= 639.9$\,m\AA) points to an age below $\sim125$\,Myr. Adding this age constraint in the BANYAN II analysis raises the THA membership probability further (P=99.93\%). In addition to a membership likelihood, the BANYAN analysis provides a kinematic distance to an average precision of $\sim$$10\%$ for young moving group members (See Figure~5 in \citealt{Malo:2013} and Figure~8 in \citealt{Gagne:2014d}).  The kinematic distance for J0219--3925 is $39.4\pm2.6$\,pc. The $7\%$ uncertainty includes both the contribution from the uncertainties on the proper motion of J0219 and the space velocity scatter of THA members.

As an additional confirmation, we used J0219--3925 apparent $I_{c}$ and $J$ magnitudes, the amplitude of the proper motion and the sky position to verify its membership probability in the BANYAN I analysis \citep{Malo:2013}. It yields an a priori probability of P=96.3\% of membership in THA. Similarly to the BANYAN\,II analysis, it is possible to include the radial velocity as an additional constraint in the analysis. When we do so, the probability is increased to P$_{\nu}$=99.9\%. The statistical distance inferred by this analysis is the same as predicted by BANYAN~II.
\label{membership}

\subsection{The Age of THA}

The Tucana and Horologium Associations were discovered independently by \citealt{Zuckerman:2000} and \citealt{Torres:2000}. Further studies confirmed that what was initially considered as two associations were more likely to constitute a single young moving group to be called the Tucana-Horologium association (THA; \citealt{Zuckerman:2001}). THA is part of the so-called Great Austral Young Association (GAYA; \citealt{Torres:2008}),  including the Columba and Carina associations. Several studies \citep{Zuckerman:2004, Torres:2008, Kiss:2011} proposed members of THA based on their galactic space velocity, galactic position, and signs of youth, yielding  a sample of members with spectral type between A1V and early-M spawning a distance range of 36 to 71\,pc. Recent studies \citep{Rodriguez:2013,Moor:2013,Malo:2013,Gagne:2014a,Gagne:2014c, Gagne:2014d,Kraus:2014} proposed more than 200 strong THA candidate members, which still require more robust kinematics to confirm their membership.

The age range of THA was estimated  to be $10-40$~Myr \citep{Zuckerman:2001, Zuckerman:2000} based on various age indicators (H$\alpha$, lithium, HR diagram). More recently, \citealt{Kraus:2014} derived an average isochronal age of 30~Myr using a new sample of 142 candidate members combined with the BCAH models \citep{Baraffe:1998}. 

In addition to the isochronal age, the lithium depletion boundary (LDB) is a key indicator to determine the age of the low-mass star population. This method was  used by \citealt{Kraus:2014} to determine the THA lithium depletion age of 41$\pm$2 and 38$\pm$2 using the BCAH and \citealt{DAntona:1997} models, respectively.

This discrepancy between isochronal and LDB ages was already demonstrated by several studies \citep{ Song:2002,Yee:2010, Binks:2014}. Recently, \citealt{Malo:2014a} have shown that using new Dartmouth Magnetic evolutionary models \citep{Feiden:2013}, the isochronal age is in better agreement with the LDB age for young low-mass stars. In general, isochronal ages are revised upward with the inclusion of magnetic fields. As shown in \citet{Malo:2014}, magnetic field strength measurements from high-resolution spectroscopy would further constrain its age. Pending such measurement, we conservatively adopt an age range of $30-40$~Myr for the THA association, and thus, for J0219--3925 system.

%The isochronal age for J0219--3925 is derived using the bolometric luminosity, effective temperature and Dartmouth Magnetic evolutionary models given magnetic field strength of 1, 2 and 2.5kG, which yields to an isochronal age of 24.7$\pm$4, 37.8$\pm$6 and 48$\pm$8Myr, respectively. 

\subsection{Spectral type and Gravity Indicators}
\label{gravity}

\begin{figure*}[!htbp]
\begin{center}
\includegraphics[width=0.99\textwidth]{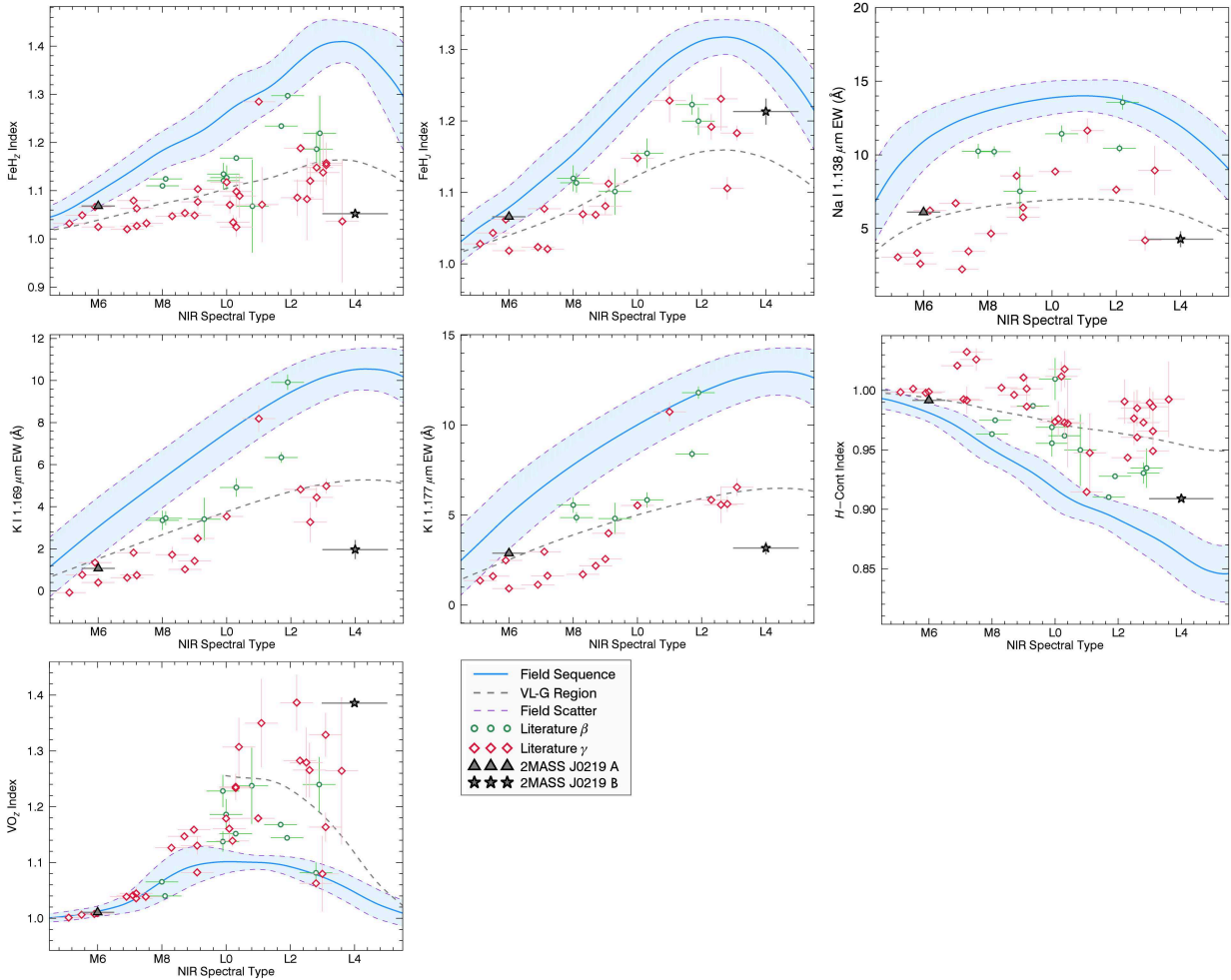}
\end{center}

\caption{\label{gravitindices}
Spectral indices as defined by \citealt{Allers:2013} for J0219--3925 and J0219--3925\,B, intermediate (INT-G or $\beta$) and very-low gravity (VL-G or $\gamma$) M and L dwarfs, the field sequence (thick, blue line) and its scatter (blue shaded region). The dotted line represents the delimitation between INT-G and VL-G regimes. Spectral types were offset by small ($<$ 0.15) random subtypes so that vertical error bars can be distinguished. For all indices, J0219--3925\,B falls outside the enveloppe of field objects, clearly highlighting its low gravity. Four indices track both metallicity and surface gravity trends (KI$_J$, FeH$_J$, $H$-cont and FeH$_z$); J0219--3925\,B does not show clear signs of peculiar metallicity, following the general trends of very-low gravity members. The M6\,$\gamma$ J0219--3925 falls in a part of the diagram where NIR low-gravity indices are less efficient than for its L4\,$\gamma$ companion. Alkali line equivalent widths are systematically lower than field objects and  consistent with   very-low gravity objects.  Individual objects were drawn from \citealt{Allers:2013} and \citealt{Manjavacas:2014}. }
\end{figure*}

We used the method of K.~Cruz et al. (in preparation; see  \citealp{Cruz:2007}) to assign a spectral type to both components of J0219--3925. This method consists of a visual comparison with a grid of spectral templates while normalizing each NIR band individually. Inspecting the slope and shape of several features in each band allows to choose a template that best matches the observations. Our sequence of field, intermediate-gravity and low-gravity templates were build by median-combining various objects within each spectral types and gravity class. The spectra used to build the templates were obtained from \cite{Allers:2013} and the SpeX Prism Spectral Libraries\footnote{\url{http://pono.ucsd.edu/\textasciitilde adam/browndwarfs/spexprism}}. Both the visual comparisons of J0219--3015 and b yielded best matches to very low gravity templates; J0219--3015 and its companion were assigned a spectral type of M6~$\gamma$ and  L4~$\gamma$ respectively. The gravity classification scheme of \cite{Allers:2013} was subsequently used to confirm that both objects have weaker alkali lines compared to field dwarfs of the same spectral types, and that J0219--3015\,B has stronger VO absorption and a triangular-shaped $H$-band continuum (which was already apparent from the visual comparison). Both objects were consistently categorized as very low gravity dwarfs by this index-based classification scheme. We show that they both display typical spectroscopic signatures of a low-gravity, including lower-than-normal alkali (\ion{K}{1}, \ion{Na}{1} and FeH) equivalent widths.  The lower gravity causes a lower pressure in the atmosphere, which decreases both the effects of pressure broadening (responsible for the lower alkali equivalent widths) and collision-induced absorption (CIA) of the H$_2$ molecule. The triangular $H$--band continuum is shaped by water absorption. In the case of field brown dwarfs, the H$_2$ CIA redirects part of the flux to the bluer side, which masks the triangular shape of the $H$-band continuum (see \citealt{Rice:2010} for further detail). Figure~\ref{gravitindices} illustrates the gravity-sensitive spectroscopic indices defined by \cite{Allers:2013} for both components of the J0219--3925 system.

The \emph{intermediate gravity} and \emph{very low gravity} classifications generally correspond to the $\beta$ and $\gamma$ gravity classifications in the optical (\citealt{Kirkpatrick:2005}; \citealt{Kirkpatrick:2006}; \citealt{Cruz:2009}), hence we choose here to use the Greek-letter nomenclature even though our gravity classification was done in the NIR.

% \cite{Allers:2013} devised a quantitative NIR classification scheme based on these spectroscopic properties to assign a gravity classification to M6--L7 dwarfs. 

\subsection{Kinematics and Common Proper Motion}
\label{commonpm}
\begin{figure}[!tbp]
\includegraphics[width=0.4\textwidth]{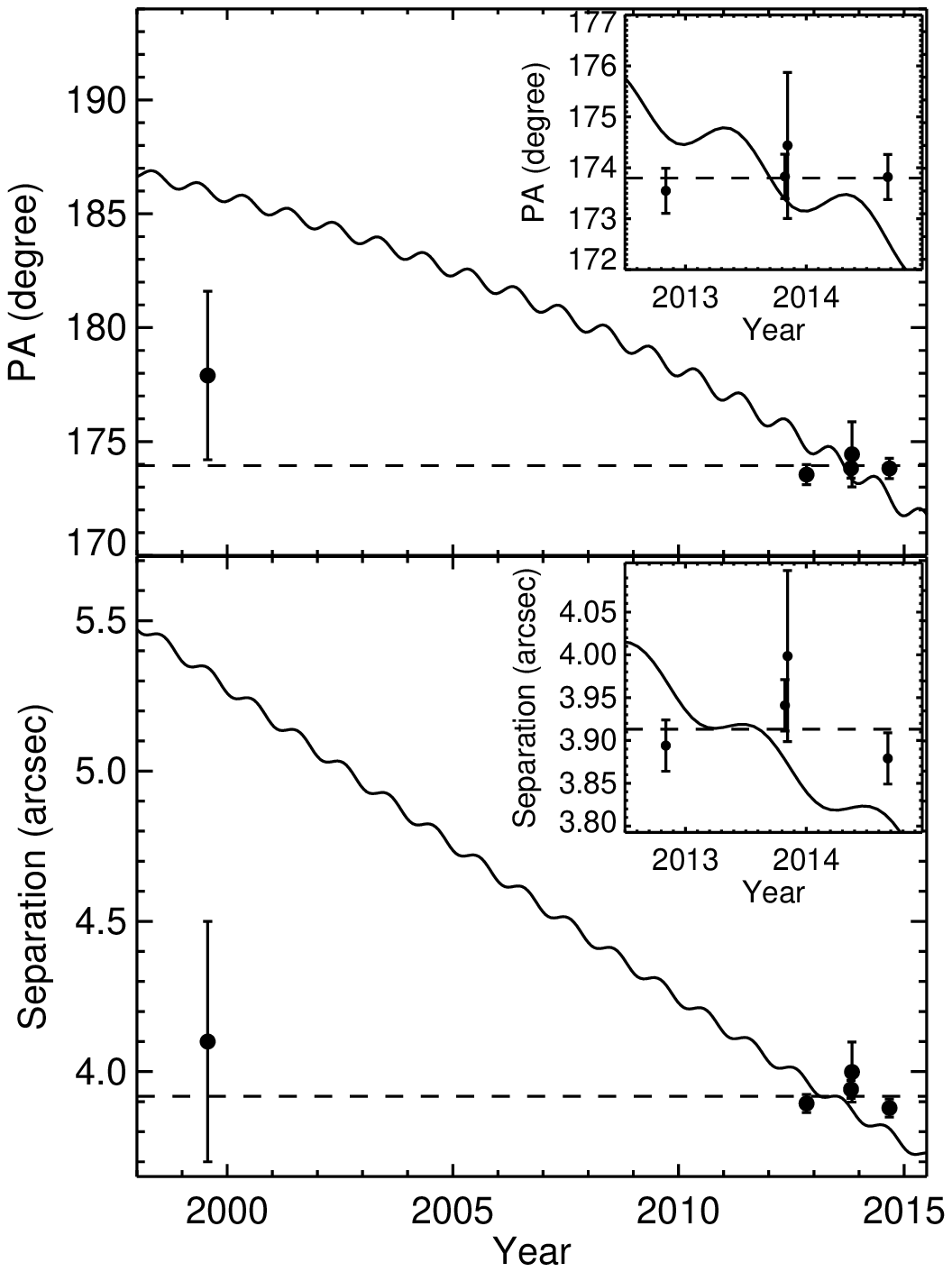}

\caption{Position angle and separation for the 2MASS (1999 July 27), GMOS $i$-band (2012 Novembre 2), F2 $J$-band  (2013 Octobre 28, 2014 Septembre 4) and SIMON $J$-band (2013 Novembre 5) imaging. PA and separation are constant to within astrometric uncertainties. Insets in each panel shows the astrometric measurements obtained since 2012. Dashed line shows the mean PA and separation, while the continuous line shows the expected values for a distant background object.  \label{comoving} }
\end{figure}

Astrometric measurements were performed on GMOS-S $i$-band, F2 $J$-band and SIMON imaging obtained on 2013 November 5. SIMON imaging obtained in early 2014 were not included as they were taken under poor seeing conditions ($\sim2\arcsec$) and do not  constrain proper motion. Archival 2MASS imaging were also used; it provides only modest constraints on the PA and separation, but the long time baseline ($\sim$15\,years) makes it a useful additions to confirm common proper motion. Astrometric errors were estimated to be at the $\sim$$0\farcs03$ level for F2 and GMOS-S data and $\sim$$0\farcs1$ for SIMON imaging. Astrometric measurements are shown in Figure~\ref{comoving}. The $\chi^2$ for the co-moving case is 5.9 for 8 degrees of freedom. The co-moving case is equivalent to a $0.9-\sigma$    event in a Gaussian distribution while the background object model would correspond to a $6.0-\sigma$   event. While this demonstrates that J0219--3925\,B is comoving to within astrometric uncertainty, any interlopers would be an L dwarf at roughly the distance of J0219--3925 with a significant proper motion that could, conceivably, match that of J0219--3925 within astrometric uncertainties. The strongest argument in favour of the two objects forming a physical pair arrises from the relative rarity of field L dwarfs per sky surface unit. 

%   print,300*(6.^2-2.^2)*!dpi/(3600.^2*41252)
Finding a low-gravity L dwarf within $4\arcsec$ of an unrelated young M star is very unlikely. Considering that we started with a sample of 300 young stars, the area within a $2-6\arcsec$ annulus around these stars covers $5.6\times 10^{-8}$ of the entire celestial sphere. With an L dwarf spatial density of $\sim$$3.8\times10^{-3}$\,pc$^{-3}$ \citep{Cruz:2007b}, there should be $\sim$$5500$ L dwarfs within 70\,pc. The likelihood of finding a coincident L dwarf to a young star in our sample is about $\sim3\times10^{-4}$ without considering the fact that the L dwarf itself also displays a low surface gravity. 

% print,4*2*!dpi/6000.  ; pm
%  print,160.^(3/2.)/(.12)^.5 
% print, (160*2*!dpi*1.5d8)/(5800.*365*86400.)
Orbital motion of the pair is expected to be close to the detection threshold. With a total mass of $\sim$$0.12$\,M$_\odot$ (see table~\ref{tblparams}) and a separation of $\sim$160\,AU, the orbital period is on the order of 6000\,yr. Assuming a face-on orbit, this leads to an orbital motion of $4$\,mas/yr. Orbital velocity is on the order of 800\,m/s and within reach {of upcoming high-resolution infrared spectrographs such as VLT/CRIRES+ \citep{Follert2014}}. Both measurements are challenging but possible with existing facilities, but not with the dataset in hand. Proper constraints on the differential motion within the system will allow one to constrain whether the orbit of J0219--3925\,B is consistent with a circular orbit or a highly eccentric one. Constraints on orbital eccentricity for a few systems similar to J0219--3925 could set strong limits on the plausible formation mechanisms. These measurements would provide constraints on the J0219--3925 system similar  to the ones obtained for a few similar companions in the planetary-mass regime by \citealt{Ginski:2014}. These authors find that fitted orbits suggest that distant companions' motion is consistent with eccentric orbits; whether this holds for the J0219--3925 system remains to be determined.

\subsection{Model Fitting with Spectroscopy and Photometry} \label{fittingmodel}

To estimate fundamental parameters of mass, effective temperature, gravity, and radius for both J0219--3925 and J0219--3925\,B we used two distinct approaches with the same set of theoretical models. First, we constrained the bulk properties of J0219--3925\,A and J0219--3925\,B (mass, effective temperature, surface gravity, radius and luminosity) by comparing the absolute magnitudes in each near-infrared bandpass with BT-Settl models\footnote{https://phoenix.ens-lyon.fr/Grids/BT-Settl/CIFIST2011/} (CIFIST 2011 opacities; \citealt{Allard:2014}) at  $30$ and $40$\,Myr. Absolute magnitudes were estimated from the kinematic distance; uncertainties in the kinematic distance were assumed to be $7$\% (see subsection~\ref{membership}), leading to a 0.14\,mag uncertainty on the absolute magnitudes.   Table~\ref{tblparams} compiles the value derived for each photometric bandpass, and the mean value for all three bandpasses is used as the best estimate. Upper and lower bounds are averaged for this best estimate, but not divided by the square root of the number of measurements as these are correlated (i.e., same age assumption, same uncertainty on distance). Overall, the mass estimate range of 12--15\,M$_{\rm Jup}$ sets the companion at the brown dwarf/planet regime limit.  Uncertainties for $\log$ g values are small (0.02 and 0.06 respectively) as the surface gravity varies very little with effective temperature for these relevant masses and the main contribution to this uncertainty arrises from the uncertainty on the age of the system. The relative uncertainty on the mass ratio (q) is smaller ($\sim6\,\%$) than the relative uncertainty on either component's mass ($\sim10\,\%$) as the derived masses for both components correlate through the distance and age estimates. 

\begin{table*}[!htb]
 \caption{Constraints on properties from models.}
\label{tblparams}
\begin{tabular}{ l cc }
\hline\hline
& J0219--3925\,A & J0219--3925\,B\\
\hline\hline

Mass (M$_J$) & $114\pm13$\,M$_{\rm Jup}$ & $13.0\pm0.7$\,M$_{\rm Jup}$\\
Mass (M$_H$) & $108\pm11$\,M$_{\rm Jup}$ & $13.9\pm1.1$\,M$_{\rm Jup}$\\
Mass (M$_K$) & $115\pm13$\,M$_{\rm Jup}$ & $14.8\pm1.6$\,M$_{\rm Jup}$\\ \hline
Mass (mean) & $113\pm12$\,M$_{\rm Jup}$  &Ê $13.9\pm1.1$\,M$_{\rm Jup}$\\
\hline\hline
q (M$_J$) & \multicolumn{2}{c}{ $0.113\pm0.007$ } \\
q (M$_H$) & \multicolumn{2}{c}{ $0.128\pm0.006$ } \\
q (M$_K$) & \multicolumn{2}{c}{ $0.127\pm0.007$ } \\ \hline
q (mean) & \multicolumn{2}{c}{  $0.123\pm0.006$} \\
\hline\hline
T$_{\rm eff}$ (M$_J$) & $3070\pm73$\,K & $1615\pm41$\,K\\
T$_{\rm eff}$ (M$_H$) & $3047\pm84$\,K & $1686\pm43$\,K\\
T$_{\rm eff}$ (M$_K$) & $3074\pm72$\,K & $1746\pm49$\,K\\ \hline
T$_{\rm eff}$ (mean)& $3064\pm76$\,K  &   $1683\pm43$\,K\\
\hline\hline
$\log$ Luminosity (M$_J$) & $-2.22\pm0.06$\,L$_{\odot}$ & $-3.92\pm0.03$\,L$_{\odot}$\\
$\log$ Luminosity (M$_H$) & $-2.26\pm0.06$\,L$_{\odot}$ & $-3.84\pm0.05$\,L$_{\odot}$\\
$\log$ Luminosity (M$_K$) & $-2.22\pm0.06$\,L$_{\odot}$ & $-3.76\pm0.06$\,L$_{\odot}$\\ \hline
$\log$ Luminosity (mean) & $-2.23\pm0.06$\,L$_{\odot}$  & $-3.84\pm0.05$\,L$_{\odot}$\\
\hline\hline
$\log g$ (M$_J$) & $4.59\pm0.06$\,cm\,s$^{-2}$ & $4.23\pm0.02$\,cm\,s$^{-2}$\\
$\log g$ (M$_H$) & $4.59\pm0.06$\,cm\,s$^{-2}$ & $4.24\pm0.04$\,cm\,s$^{-2}$\\
$\log g$ (M$_K$) & $4.59\pm0.06$\,cm\,s$^{-2}$ & $4.25\pm0.06$\,cm\,s$^{-2}$\\ \hline
$\log g$ (mean) Ê&Ê$4.59\pm0.06$\,cm\,s$^{-2}$& $4.24\pm0.04$\,cm\,s$^{-2}$\\
\hline\hline
Deuterium (M$_J$) & 0 & $0.77\pm0.06$\\
Deuterium (M$_H$) & 0 & $0.68\pm0.07$\\
Deuterium (M$_K$) & 0 & $0.61\pm0.09$\\ \hline
Deuterium (mean) Ê&Ê  0 &  $0.69\pm0.07$\\
\hline\hline
Radius (M$_J$) & $2.75\pm0.14$\,R$_{\rm Jup}$ & $1.41\pm0.02$\,R$_{\rm Jup}$\\
Radius (M$_H$) & $2.68\pm0.12$\,R$_{\rm Jup}$ & $1.44\pm0.04$\,R$_{\rm Jup}$\\
Radius (M$_K$) & $2.76\pm0.14$\,R$_{\rm Jup}$ & $1.46\pm0.04$\,R$_{\rm Jup}$\\ \hline
Radius (mean)  & $2.73\pm0.13$\,R$_{\rm Jup}$ &  $1.44\pm0.03$\,R$_{\rm Jup}$\\
\hline\hline\\

\end{tabular}
\end{table*}

\begin{figure*}[!th]
\includegraphics[width=0.95\textwidth]{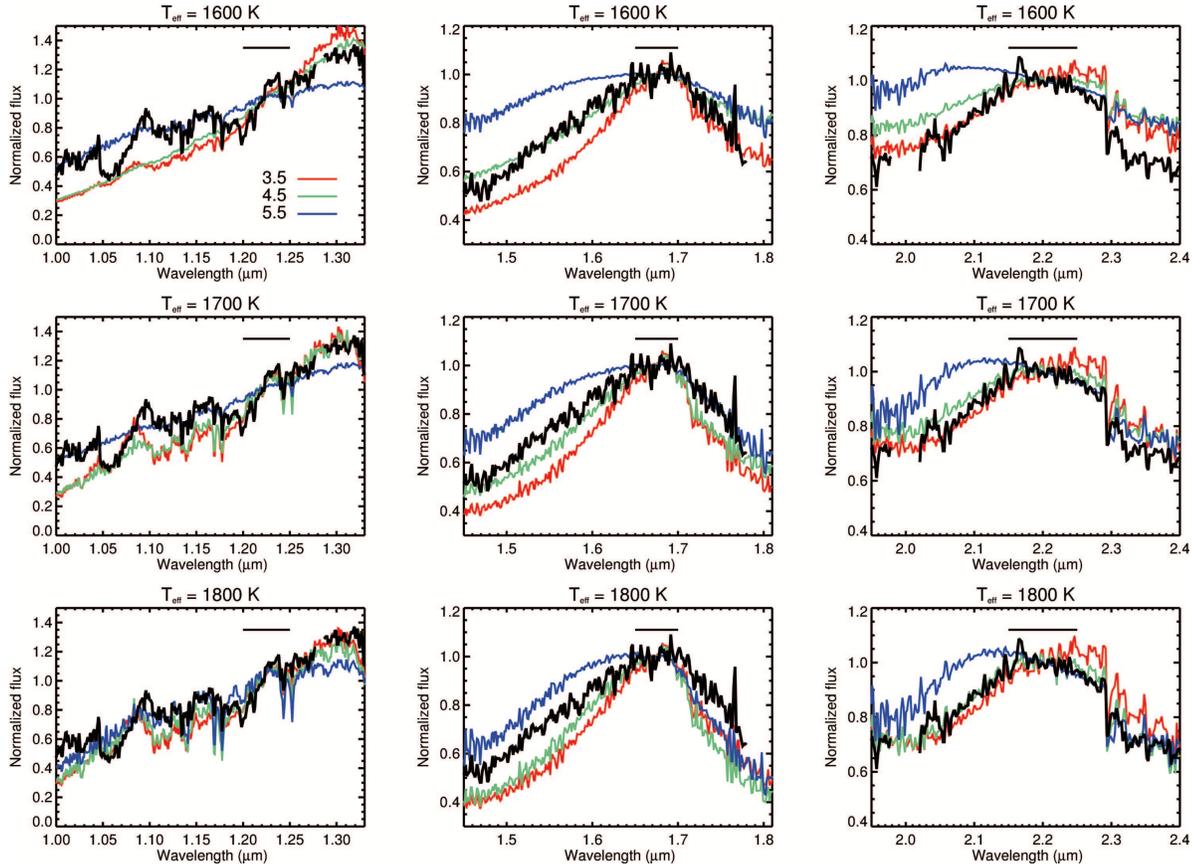}
\caption{
{Comparison between 2M0219--3925\,B  (black) and BT-Settl models (red, green and blue) in the $Y+J$, $H$ and $K$ spectral regions. To highlight shape differences  in SED of individual photometric bandpasses, we show all models normalized to a common wavelength interval (shown as a thick horizontal line in each plot). For the sake of clarity, only $\log g$ values of 3.5, 4.5 and 5.5 and temperatures from $T_{\rm eff}=1600$\,K to 1800\,K are plotted.
}
\label{spfit}}
\end{figure*}

The second approach to constrain the properties of J0219--3925\,B has been to perform a $\chi^2$ fit between  the observed and theoretical spectra in 3 individual bandpasses ($Y+J$, $H$ and $K_s$; respectively $1.00-1.33$\,$\mu$m, $1.45-1.81$\,$\mu$m and $1.95-2.40\mu$m). For each bandpass, spectra were normalized over part of the domain (horizontal lines in Figure~\ref{spfit}). By performing this normalization, we assume no prior knowledge of distance and absolute magnitude. The spectral fitting analysis has been performed by steps of 0.5\,dex in $\log g$ between 3.5 and 5.5, and over the 1500\,K to 1800\,K domain. Models at intermediate gravity (4.5) better fit the SED. Within the $H$ band (middle columns), the higher-gravity (5.5) models show the distinctive flattening of the SED, which is the hallmark of old, field, L dwarfs. Within the $K$ band (right), there is a clear shift of the SED peak from $\sim2.1$\,$\mu$m to $\sim2.25$\,$\mu$m between $\log g=5.5$ and $\log g=3.5$, for all temperatures, with 2M0219--3925\,B being intermediate between these scenarios. For $Y+J$ and $K$, the best-fitting model has a temperature of 1700\,K and $\log g=4.5$. The value derived for $H$ is only marginally different with $T_{\rm eff}=1600$\,K and $\log g=4.0$. These values are in very good agreement with those derived from photometry alone (see Table~\ref{tblparams}), with $T_{\rm eff}=1683\pm43$\,K and $\log g=4.24{\pm0.04}$.

 Overall, the mass of J0219--3925\,B falls squarely at the upper limit of the International Astronomical Union definition for an exoplanet -- {\it consisting of an object with a true mass below the limiting mass for thermonuclear fusion of deuterium} -- and is in orbit around a star, indeed J0219--3925 has a mass estimate above the hydrogen burning limit for plausible ages of THA. Evolution models predict that J0219--3925\,B will have retained $\sim$70\,$\%$ of its initial deuterium by 40\,Myr.  As it is expected to have partially burned its deuterium, and considering its brightness, it would constitue a good target to spectroscopically test the brown dwarf/planet boundary, that is expected to be moderately metallicity-dependent \citep{Spiegel:2011} and its exact predicted location varies by $\sim$1\,$M_{\rm Jup}$ depending on atmosphere models \citep{Saumon:2008}. The exact nomenclature for such an object is still to be properly defined, and it joins a number of other objects such as J0103--5515(AB)b and J0122--2439\,B \citep{Delorme:2013, Bowler:2013a} at the upper-limit of planethood.  As the J0219--3925\,B is expected to have burned some of its deuterium and formally lies just above the deuterium-burning limit, we use the "B" designation for stellar and brown dwarf companions rather than the "b" in usage for planetary companions even though it is predicted to be slightly less massive than objects initially described as planetary companions (e.g., AB Pictoris\,b).

\section{DISCUSSION}
\label{discussion}

\begin{figure*}[!th]

\includegraphics[width=0.95\textwidth]{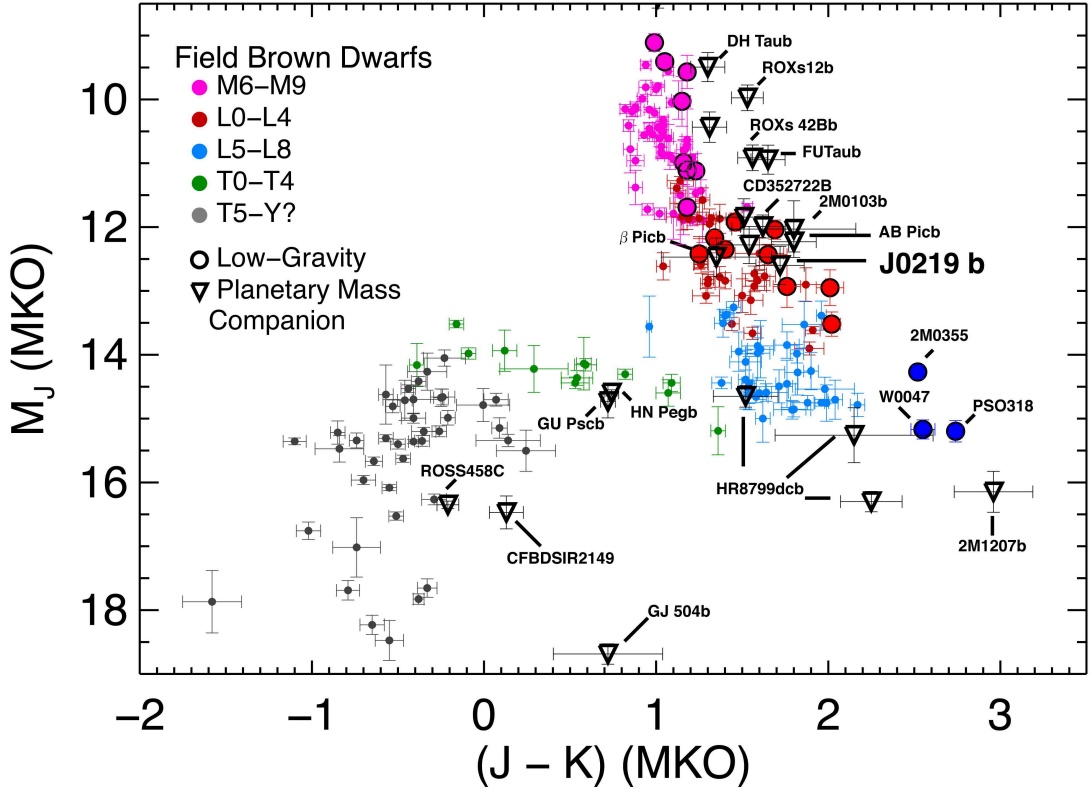}
\caption{Color-magnitude plot of field dwarfs, low-gravity and directly imaged companions. J0219--3925\,B is redder than typical field L dwarfs, and falls within the sequence of planetary-mass companions.  \label{colmag} }
\end{figure*}

Figure~\ref{colmag} illustrates the position of J0219--3925\,B in the $J-K$ color-magnitude diagram compared to field ultracool dwarfs and planetary mass objects. J0219--3925\,B follows the overall trend of planetary-mass companions in being redder by about 0.5\,mag than field ultracool dwarfs of similar $M_K$.

With a mass ratio $q=0.122\pm0.006$, the J0219--3925 system is similar to a heavier version of  2MASSW J1207334-393254 \citep{Chauvin:2005a}, consisting of a $4-6$\,$M_{\rm Jup}$ planet around a $25-45$\,$M_{\rm Jup}$ brown dwarf (mass ratio $0.13-0.16$). With a separation of $>55$\,AU, J1207-3932 system comes very close to J0219--3925 in terms of binding energy, with a total mass $\sim$3 times larger ($\sim40$\,M$_{\rm Jup}$ versus $\sim120$\,$M_{\rm Jup}$), but a physical separation that is also about three times larger ($>$$55$ versus $>$$160$\,AU). It also comes close to the J0103-5515(AB)b system, which consists of a $12-14$\,M$_{\rm Jup}$ companion to a binary mid-M \citep{Delorme:2013}. We can rule-out the possibility that J0219--3925  is a triple analogous to J0103-5515(AB)b. We only have seeing-limited observations of the system, so a hierarchical system with an inner component tighter than $\sim$$0\farcs5$ would not be resolved with the dataset at hands. The best constraints on the presence of an inner binary comes from the $M_J$, $M_H$ and $M_{K_s}$ derived from absolute magnitude versus spectral type relations for young field dwarfs. These relations predict contrast ratios respectively of 4.36, 3.78, 3.31\,mag (Gagn\'e et al., submitted to ApJ) for  single M6\,$\gamma$ and L4\,$\gamma$ components, and these values are in close agreement with the observed contrasts of $\Delta J=4.2\pm0.1$,  $\Delta H=3.8\pm0.1$ and $\Delta {K_s}=3.4\pm0.1$. We can therefore conclude that neither J0219--3925\,A nor  J0219--3925\,B is an equal luminosity binary. We cannot rule-out that either object is itself a high-contrast ($>1$\,mag) binary.

J0219--3925\,B also shares various similarities with  AB Pictoris\,b \citep{Chauvin:2005}. Both objects are companions to Tucana-Horologium members and therefore share a common age. AB Pictoris\,b as a slightly earlier spectral type (L$0.5\pm0.5$) and $\sim$$0.4$\,mag brighter in both $J$ and $K_s$ \citep{Biller:2013b}.  The brightness difference between the two is consistent with what would be expect considering their respective spectral types  (see Figure\,17 in \citealt{Biller:2013b}). As shown in Figure~\ref{sp1}, right panel, J0219--3925\,B displays slightly deeper water bands, especially around 1.5\,$\mu$m and 1.95\,$\mu$m and a deeper CO bandhead longward of 2.29\,$\mu$m, consistent with a  later spectral type. This brightness difference corresponds to a mass difference of $\sim$1\,M$_{\rm Jup}$ between the two objects and a comparative study could be used to better constrain the deuterium burning limit through high resolution spectroscopy.

\begin{figure*}[!th]

\includegraphics[width=0.95\textwidth]{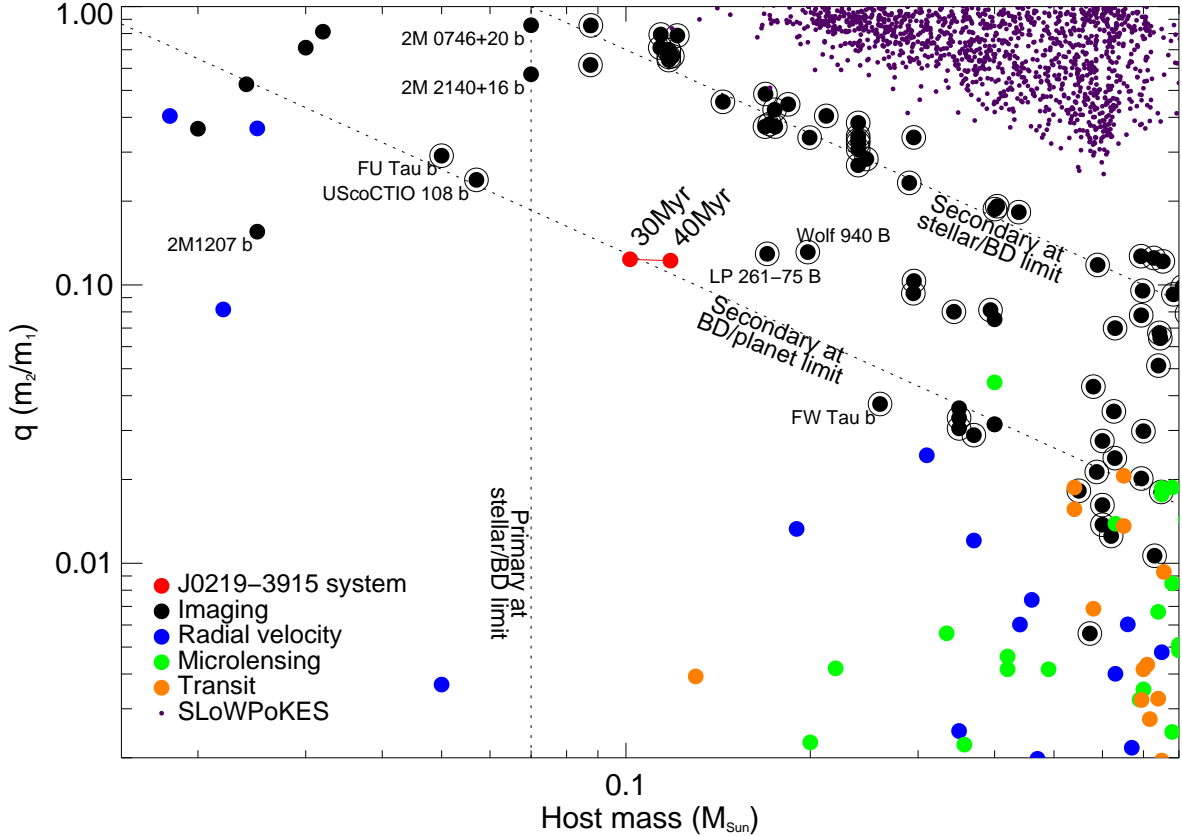}
\caption{ Host mass versus mass ratio for substellar companions detected through direct imaging (black), radial-velocity (blue), microlensing (green) or transit (orange). Also included is the very large sample of binaries including an cool dwarf from the SLoWPoKES survey \citep{Dhital2010}. Imaged binaries with orbital separations larger than 100\,AU are circled. Dashed lines indicate the notional 13\,M$_{\rm Jup}$ BD/planet limit for companions and the brown dwarf/stellar limit for both hosts and companions and selected systems are labelled. Some hosts (e.g., 2M 0746+20; \citealt{Konopacky:2010}) are themselves near-equal mass binaries; the total mass of the central binary is given. The position of the J0219--3925 system for ages of 30 and 40\,Myr is shown. The J0219--3925 system falls in an empty part of the diagram, intermediate in properties between near-equal mass binaries such as 2M\,0746+20 and 2M\,2140+16 \citep{Konopacky:2010}  and planets uncovered through radial velocity measurements and imaging of young systems. Most systems close to the stellar/substellar limit are L dwarfs companions to field stars found through wide-field surveys. System parameters were drawn from the Extrasolar Planets Encyclopedia, \citealt{Deacon2014}, \citealt{Baron2015} and \citealt{Konopacky:2010}.  }
 \label{massvsq}
 \end{figure*}

Figure~\ref{massvsq} shows the host mass versus mass ratio for imaged planetary and low-mass brown dwarf companions. Interestingly, the J0219--3925 system falls in a relative gap among known systems in that diagram. Known systems hosting a low-mass secondary ($<40\,$$M_{\rm Jup}$) either have near-equal mass; many of these systems have been identified as near-equal luminosity field brown dwarfs or have a much smaller $q$ values (typically $q<2$\%) and have been uncovered with other observing techniques. Efficient searches for near-equal luminosity sub-stellar binaries have been largely performed with HST (e.g., \citealt{Burgasser2006b}) and laser guide star adaptive-optics imaging (e.g., \citealt{Liu2012}) while low-q binaries have largely been identified by other techniques such as microlensing, transit or radial velocity.   Wide-field imaging surveys have also uncovered a few tens of wide companions to field stars (e.g., \citealt{Baron2015}, \citealt{Deacon2014} and references compiled therein); many of these have mass ratios comparable to that of  J0219--3925, but with significantly more massive components. The two companions uncovered that come closest to J0219--3925 are LP\,261-75 and Wolf 940. LP\,261-75\,AB consists of a young M4.5/L6 binary at a projected separation of 450\,AU \citep{Burgasser2005, Reid2006}.  LP\,261-75 does not match the kinematic properties of young moving groups included in the BANYAN~II tool, but its primary has chromospheric activity levels consistent with an age of 100-200\,Myr. This youth of the system leads to a low mass estimate for the secondary of $15-30$\,M$_{\rm Jup}$. Wolf 940\,AB consists of an M4+T8.5 binary with a projected separation of $400$\,AU. The system has an age of 3.5-6\,Gyr as derived from the chromospheric activity of the host star, leading to a mass estimate of 20-32\,M$_{\rm jup}$, making it a close match the LP\,261-75 system albeit at an older age.

While pairs similar to J0219--3925 may be inherently rare, a relatively straightforward observation bias could explain the fact that similar systems have been overlooked. Hosts with spectral types as late as that of J0219--3925\,A are in general too faint at optical wavelengths to close an adaptive-optics loop on, and are therefore under-represented in planet-search surveys. Furthermore, until very recently, very few nearby ($<50$\,pc) young stars that late were known, and most comparable objects had been found in significantly more distant open clusters, making the search of faint companions even more challenging. While companions similar to J0219--3925\,B may exist around very nearby M dwarfs, the contrast ratio increases significantly with age. At an age of 1\,Gyr the pair would have a contrast ratio of $\Delta J\sim8$ and an effective temperature of $\sim600$\,K while at 5\,Gyr, it will have a contrast of $\Delta J\sim10$ and an effective temperature of $\sim370$\,K \citep{Beichman:2010}. At these magnitudes and temperatures, J0219--3925\,B will respectively have spectral types of T8.5 and Y0 \citep{Dupuy2012, Marsh:2013}. Such companions at separations of a few arcseconds around nearby mid-M dwarfs could easily have been overlooked by current brown dwarf searches using WISE \citep{Wright:2010} or seeing-limited near-infrared observations due to the far wings of the central star's point-spread function. GAIA\footnote{http://sci.esa.int/gaia/} astrometric measurements will not constrain the occurence of similar systems in the solar neighbourhood; for a face-on circular orbit and a distance of 10\,pc, the host star in a system similar to J0219--3925 will display an astrometric acceleration of $\sim1.8$\,$\mu$as/yr$^2$, below the detection threshold for this mission. The GAIA mission may nevertheless uncover similar systems in young associations through their common proper motion. The best prospect to find Gyr-old siblings of the J0219--3925 system is with the use of very deep near-infrared imaging, either under good seeing and proper point-spread function subtraction, or with laser-guide star adaptive optics.

Weak constraints can be set on the occurence of companions similar to J0219--3925\,B around M dwarfs for the range of separations explored here ($100-5000$\,AU). The PALMS survey \citep{Bowler:2015} provides the largest sample to assess the abundance of analogs to J0219--3925\,B, although the separation range probed with their adaptive optics observations only partially overlaps with the range of interest here. They set a $<10$\% upper limit on the occurence of planets between 6 and 200\,AU and $<50$\% limit for $1.8-570$\,AU (COND atmosphere models, circular orbits and 95\% confidence level). The discovery of a single object at the planetary/brown dwarf limit in our sample is therefore well within the constraints set by the PALMS survey. Our survey included stars with different levels of confirmation regarding their age drawn from different sources, so deriving clean statistical constraints on the abundance of companions similar to J0219--3925\,B is non-trivial and beyond the scope of this paper. J0219--3925 was drawn from the BASS sample of late-type objects ($>$M4). We observed 67 high-probability BASS candidates and 13 low-probability ones (see details \citealt{Gagne:2015}); the absence of any further candidate in the $2-6\arcsec$ separation range suggests that objects similar to J0219--3925\,B are relatively rare around M dwarfs, with an occurence rate of $<6.8\%$ at the $95\%$ confidence level.

\acknowledgments

\textit{Acknowledgments}

The authors thank Noel Richardson for thoughtful discussions regarding the characterization of the J0219--3925 system. This paper includes data gathered with the 6.5 meter Magellan Telescopes located at Las Campanas Observatory, Chile. Based on observations obtained at the Gemini Observatory, which is operated by the  Association of Universities for Research in Astronomy, Inc., under a cooperative agreement  with the NSF on behalf of the Gemini partnership: the National Science Foundation  (United States), the National Research Council (Canada), CONICYT (Chile), the Australian  Research Council (Australia), Minist\'{e}rio da Ci\^{e}ncia, Tecnologia e Inova\c{c}\~{a}o  (Brazil) and Ministerio de Ciencia, Tecnolog\'{i}a e Innovaci\'{o}n Productiva (Argentina). This publication makes use of data products from the Two Micron All Sky Survey, which is a joint project of the University of Massachusetts and the Infrared Processing and Analysis Center/California Institute of Technology, funded by the National Aeronautics and Space Administration and the National Science Foundation. The Digitized Sky Surveys were produced at the Space Telescope Science Institute under U.S. Government grant NAG W-2166. The images of these surveys are based on photographic data obtained using the Oschin Schmidt Telescope on Palomar Mountain and the UK Schmidt Telescope. The plates were processed into the present compressed digital form with the permission of these institutions.

\

\expandafter\ifx\csname natexlab\endcsname\relax\def\natexlab#1{#1}\fi


\begin{thebibliography}{63}
\expandafter\ifx\csname natexlab\endcsname\relax\def\natexlab#1{#1}\fi

%%%%%%%%%%%%%%%%%%%%%%%%%%%%%%%%%%%%%%%%


\bibitem[{{Alibert} {et~al.}(2005)}]{Alibert2005}
{Alibert}, Y., {Mordasini}, C., {Benz}, W. \& {Winisdoerffer}, C. 2005, \aap, 434, 343

%%%%%%%%%%%%%%%%%%%%%%%%%%%%%%%%%%%%%%%%

\bibitem[{{Allard}(2014)}]{Allard:2014}
{Allard}, F. 2014, in IAU Symposium, Vol. 299, IAU Symposium, ed. M.~{Booth},
  B.~C. {Matthews}, \& J.~R. {Graham}, 271--272

\bibitem[{{Allers} \& {Liu}(2013)}]{Allers:2013}
{Allers}, K.~N., \& {Liu}, M.~C. 2013, \apj, 772, 79

\bibitem[{{Baraffe} {et~al.}(1998){Baraffe}, {Chabrier}, {Allard}, \&
  {Hauschildt}}]{Baraffe:1998}
{Baraffe}, I., {Chabrier}, G., {Allard}, F., \& {Hauschildt}, P.~H. 1998, \aap,
  337, 403


\bibitem[{{Baron} {et~al.}(2015)}]{Baron2015}
{Baron}, F., {et~al.} 2015, \apj, 802, 37

\bibitem[{{Beichman} {et~al.}(2010){Beichman}, {Krist}, {Trauger}, {Greene},
  {Oppenheimer}, {Sivaramakrishnan}, {Doyon}, {Boccaletti}, {Barman}, \&
  {Rieke}}]{Beichman:2010}
{Beichman}, C.~A., {et~al.} 2010, \pasp, 122, 162

\bibitem[{{Biller} {et~al.}(2013){Biller}, {Liu}, {Wahhaj}, {Nielsen},
  {Hayward}, {Males}, {Skemer}, {Close}, {Chun}, {Ftaclas}, {Clarke}, {Thatte},
  {Shkolnik}, {Reid}, {Hartung}, {Boss}, {Lin}, {Alencar}, {de Gouveia Dal
  Pino}, {Gregorio-Hetem}, \& {Toomey}}]{Biller:2013b}
{Biller}, B.~A., {et~al.} 2013, \apj, 777, 160

\bibitem[{{Binks} \& {Jeffries}(2014)}]{Binks:2014}
{Binks}, A.~S., \& {Jeffries}, R.~D. 2014, \mnras, 438, L11

\bibitem[{{Bonnefoy} {et~al.}(2014){Bonnefoy}, {Chauvin}, {Lagrange}, {Rojo},
  {Allard}, {Pinte}, {Dumas}, \& {Homeier}}]{Bonnefoy:2014}
{Bonnefoy}, M., {Chauvin}, G., {Lagrange}, A.-M., {Rojo}, P., {Allard}, F.,
  {Pinte}, C., {Dumas}, C., \& {Homeier}, D. 2014, \aap, 562, A127

%%%%%%%%%%%%%%%%%%%%%%%%%%%%%%%%%%%%%%%%%
%Boss1997

\bibitem[{{Boss} (1997){Boss}}]{Boss1997}
{Boss}, A.~P. 1997, Science, 276, 1836


%%%%%%%%%%%%%%%%%%%%%%%%%%%%%%%%%%%%%%%%%

\bibitem[{{Bowler} {et~al.}(2013){Bowler}, {Liu}, {Shkolnik}, \&
  {Dupuy}}]{Bowler:2013a}
{Bowler}, B.~P., {Liu}, M.~C., {Shkolnik}, E.~L., \& {Dupuy}, T.~J. 2013, \apj,
  774, 55

\bibitem[{{Bowler} {et~al.}(2015){Bowler}, {Liu}, {Shkolnik}, \&
  {Tamura}}]{Bowler:2015}
{Bowler}, B.~P., {Liu}, M.~C., {Shkolnik}, E.~L., \& {Tamura}, M. 2015, \apjs,
  216, 7

\bibitem[{{Brandt} {et~al.}(2014){Brandt}, {McElwain}, {Turner}, {Mede},
  {Spiegel}, {Kuzuhara}, {Schlieder}, {Wisniewski}, {Abe}, {Biller},
  {Brandner}, {Carson}, {Currie}, {Egner}, {Feldt}, {Golota}, {Goto}, {Grady},
  {Guyon}, {Hashimoto}, {Hayano}, {Hayashi}, {Hayashi}, {Henning}, {Hodapp},
  {Inutsuka}, {Ishii}, {Iye}, {Janson}, {Kandori}, {Knapp}, {Kudo}, {Kusakabe},
  {Kwon}, {Matsuo}, {Miyama}, {Morino}, {Moro-Mart{\'{\i}}n}, {Nishimura},
  {Pyo}, {Serabyn}, {Suto}, {Suzuki}, {Takami}, {Takato}, {Terada}, {Thalmann},
  {Tomono}, {Watanabe}, {Yamada}, {Takami}, {Usuda}, \&
  {Tamura}}]{Brandt:2014a}
{Brandt}, T.~D., {et~al.} 2014, \apj, 794, 159

\bibitem[{{Burgasser} {et~al.}(2005){Burgasser}, {Kirkpatrick} \& {Lowrance}}]{Burgasser2005}
{Burgasser}, A.~J.~B., {Kirkpatrick}, J.~D. \& {Lowrance}, P.~J. 2005, \apj, 129, 2849


\bibitem[{{Burgasser} {et~al.}(2006){Burgasser}, {Kirkpatrick}, {Cruz}, {Reid}, {Leggett}, {Liebert}, {Burrows}, \& {Brown}}]{Burgasser2006b}
{Burgasser}, A.~J.~B., {et~al.} 2011, \apjs, 166, 585

\bibitem[{{Burningham} {et~al.}(2009){Burningham}, {Pinfield}, {Leggett}, {Tinney}, {Liu}, {Homeier}, {West}, {Day-Jones}, {Huelamo},  {Dupuy}, {Zhang}, {Murray}, {Lodieu}, {Barrado Y Navascu{\'e}s}, {Folkes}, {Galvez-Ortiz}, {Jones}, {Lucas},  {Calderon} \& {Tamura}  }]{Burningham2009}
{Burningham}, B., {et~al.} 2011, \mnras, 395, 1237


\bibitem[{{Burningham} {et~al.}(2011){Burningham}, {Leggett}, {Homeier},
  {Saumon}, {Lucas}, {Pinfield}, {Tinney}, {Allard}, {Marley}, {Jones},
  {Murray}, {Ishii}, {Day-Jones}, {Gomes}, \& {Zhang}}]{Burningham2011}
{Burningham}, B., {et~al.} 2011, \mnras, 414, 3590
%%%%%%%%%%%%%%%%%%%%%%%%%%%%%%%%%%%%%%%%%%%%%%%%%%%%

 
\bibitem[{{Cameron}(1978){Cameron}}]{Cameron1978}
{Cameron}, A.~G.~W. 1978, M\&P,  18, 5

%%%%%%%%%%%%%%%%%%%%%%%%%%%%%%%%%%%%%%%%%%%%%%%%%%%%



\bibitem[{{Chauvin} {et~al.}(2005{\natexlab{a}}){Chauvin}, {Lagrange}, {Dumas},
  {Zuckerman}, {Mouillet}, {Song}, {Beuzit}, \& {Lowrance}}]{Chauvin:2005a}
{Chauvin}, G., {Lagrange}, A.-M., {Dumas}, C., {Zuckerman}, B., {Mouillet}, D.,
  {Song}, I., {Beuzit}, J.-L., \& {Lowrance}, P. 2005{\natexlab{a}}, \aap, 438,
  L25

\bibitem[{{Chauvin} {et~al.}(2005{\natexlab{b}}){Chauvin}, {Lagrange},
  {Zuckerman}, {Dumas}, {Mouillet}, {Song}, {Beuzit}, {Lowrance}, \&
  {Bessell}}]{Chauvin:2005}
{Chauvin}, G., {et~al.} 2005{\natexlab{b}}, \aap, 438, L29

\bibitem[{{Cruz} \& {N{\'u}{\~n}ez}(2007)}]{Cruz:2007}
{Cruz}, K., \& {N{\'u}{\~n}ez}, A. 2007, in Cool Stars 17, Poster Presentation

\bibitem[{{Cruz} {et~al.}(2009){Cruz}, {Kirkpatrick}, \&
  {Burgasser}}]{Cruz:2009}
{Cruz}, K.~L., {Kirkpatrick}, J.~D., \& {Burgasser}, A.~J. 2009, \aj, 137, 3345

\bibitem[{{Cruz} {et~al.}(2007){Cruz}, {Reid}, {Kirkpatrick}, {Burgasser},
  {Liebert}, {Solomon}, {Schmidt}, {Allen}, {Hawley}, \& {Covey}}]{Cruz:2007b}
{Cruz}, K.~L., {et~al.} 2007, \aj, 133, 439

\bibitem[{{Cutri} {et~al.}(2013){Cutri}, {et al.}}]{Cutri:2013}
{Cutri}, R.~M., {et~al.} 2013, VizieR Online Data Catalog, 2328, 0


\bibitem[{{D'Antona} \& {Mazzitelli}(1997)}]{DAntona:1997}
{D'Antona}, F., \& {Mazzitelli}, I. 1997, \memsai, 68, 807



\bibitem[{{Deacon} {et~al.}(2014)}]{Deacon2014}
{Deacon}, N.~R., {et~al.} 2014, \apj, 792, 119



\bibitem[{{Delorme} {et~al.}(2013){Delorme}, {Gagn{\'e}}, {Girard}, {Lagrange},
  {Chauvin}, {Naud}, {Lafreni{\`e}re}, {Doyon}, {Riedel}, {Bonnefoy}, \&
  {Malo}}]{Delorme:2013}
{Delorme}, P., {et~al.} 2013, \aap, 553, L5

\bibitem[{{DENIS Consortium}(2005){DENIS Consortium.}}]{DENIS:2005}
DENIS Consortium. 2005, yCat, 2263, 0


\bibitem[{{Dhital} {et~al.}(2010)}]{Dhital2010}
{Dhital}, S., {West}, A.~A., {Stassun}, K.~G. \& {Bochanski}, J.~J. 2010, \apj, 139, 2566


\bibitem[{{Doyon} {et~al.}(2000){Doyon}, {Nadeau}, \&
  {Vall{\'e}e}}]{Doyon:2000b}
{Doyon}, R., {Nadeau}, D., \& {Vall{\'e}e}, P. 2000, in Astronomical Society of
  the Pacific Conference Series, Vol. 195, Imaging the Universe in Three
  Dimensions, ed. W.~{van Breugel} \& J.~{Bland-Hawthorn}, 548

\bibitem[{{Dupuy} \& {Liu}(2012)}]{Dupuy2012}
{Dupuy}, T.~J., \& {Liu}, M.~C. 2012, \apjs, 201, 19

\bibitem[{{Epchtein} {et~al.}(1999){Epchtein}, {Deul}, {Derriere},
  {Borsenberger}, {Egret}, {Simon}, {Alard}, {Balazs}, {de Batz}, {Cioni},
  {Copet}, {Dennefeld}, {Forveille}, {Fouque}, {Garzon}, {Habing}, {Holl},
  {Hron}, {Kimeswenger}, {Lacombe}, {Le Bertre}, {Loup}, {Mamon}, {Omont},
  {Paturel}, {Persi}, {Robin}, {Rouan}, {Tiphene}, {Vauglin}, \&
  {Wagner}}]{Epchtein:1999}
{Epchtein}, N., {et~al.} 1999, VizieR Online Data Catalog, 2240, 0

\bibitem[{{Feiden} \& {Chaboyer}(2013)}]{Feiden:2013}
{Feiden}, G.~A., \& {Chaboyer}, B. 2013, \apj, 779, 183


\bibitem[{{Follert} {et~al.}(2014){Follert}, {Dorn}, {Oliva}, {Lizon}, {Hatzes}, {Piskunov}, {Reiners}, {Seemann}, {Stempels}, {Heiter}, {Marquart}, {Lockhart}, {Anglada-Escude}, {L{\"o}winger}, {Baade}, {Grunhut}, {Bristow}, {Klein}, {Jung}, {Ives}, {Kerber}, {Pozna}, {Paufique}, {Kaeufl}, {Origlia}, {Valenti}, {Gojak}, {Hilker}, {Pasquini}, {Smette} \& {Smoker}
}]{Follert2014}
 Follert, R., {et~al.} 2015, Proc. SPIE, 9147, 19


\bibitem[{{Gagn{\'e}} {et~al.}(2014{\natexlab{a}}){Gagn{\'e}}, {Faherty},
  {Cruz}, {Lafreni{\`e}re}, {Doyon}, {Malo}, \& {Artigau}}]{Gagne:2014a}
{Gagn{\'e}}, J., {Faherty}, J.~K., {Cruz}, K., {Lafreni{\`e}re}, D., {Doyon},
  R., {Malo}, L., \& {Artigau}, {\'E}. 2014{\natexlab{a}}, \apjl, 785, L14

\bibitem[{{Gagn{\'e}} {et~al.}(2014{\natexlab{b}}){Gagn{\'e}},
  {Lafreni{\`e}re}, {Doyon}, {Artigau}, {Malo}, {Robert}, \&
  {Nadeau}}]{Gagne:2014c}
{Gagn{\'e}}, J., {Lafreni{\`e}re}, D., {Doyon}, R., {Artigau}, {\'E}., {Malo},
  L., {Robert}, J., \& {Nadeau}, D. 2014{\natexlab{b}}, \apjl, 792, L17

\bibitem[{{Gagn{\'e}} {et~al.}(2014{\natexlab{c}}){Gagn{\'e}},
  {Lafreni{\`e}re}, {Doyon}, {Malo}, \& {Artigau}}]{Gagne:2014d}
{Gagn{\'e}}, J., {Lafreni{\`e}re}, D., {Doyon}, R., {Malo}, L., \& {Artigau},
  {\'E}. 2014{\natexlab{c}}, \apj, 783, 121

\bibitem[{{Gagn{\'e}} {et~al.}(2015){Gagn{\'e}}, {Lafreni{\`e}re}, {Doyon},
  {Malo}, \& {Artigau}}]{Gagne:2015}
---. 2015, \apj, 798, 73

\bibitem[{{Ginski} {et~al.}(2014){Ginski}, {Schmidt}, {Mugrauer},
  {Neuh{\"a}user}, {Vogt}, {Errmann}, \& {Berndt}}]{Ginski:2014}
{Ginski}, C., {Schmidt}, T.~O.~B., {Mugrauer}, M., {Neuh{\"a}user}, R., {Vogt},
  N., {Errmann}, R., \& {Berndt}, A. 2014, \mnras, 444, 2280


\bibitem[{{Girard} {et~al.}(2011){Girard}, {van Altena}, {Zacharias}, {Vieira}, {Casetti-Dinescu}, {Castillo}, {Herrera}, {Lee}, {Beers}, {Monet} \& {L{\'o}pez}}]{Girard2011}
{Girard}, T.~M., {et~al.}  2011, \aj, 142, 15


\bibitem[{{Goldman} {et~al.}(2010){Goldman}, {Marsat}, {Henning}, {Clemens}, \&
  {Greiner}}]{Goldman:2010}
{Goldman}, B., {Marsat}, S., {Henning}, T., {Clemens}, C., \& {Greiner}, J.
  2010, \mnras, 405, 1140

\bibitem[{{Kalas} {et~al.}(2008){Kalas}, {Graham}, {Chiang}, {Fitzgerald},
  {Clampin}, {Kite}, {Stapelfeldt}, {Marois}, \& {Krist}}]{Kalas2008}
{Kalas}, P., {et~al.} 2008, Science, 322, 1345

\bibitem[{{Kirkpatrick}(2005)}]{Kirkpatrick:2005}
{Kirkpatrick}, J.~D. 2005, \araa, 43, 195

\bibitem[{{Kirkpatrick} {et~al.}(2006){Kirkpatrick}, {Barman}, {Burgasser},
  {McGovern}, {McLean}, {Tinney}, \& {Lowrance}}]{Kirkpatrick:2006}
{Kirkpatrick}, J.~D., {Barman}, T.~S., {Burgasser}, A.~J., {McGovern}, M.~R.,
  {McLean}, I.~S., {Tinney}, C.~G., \& {Lowrance}, P.~J. 2006, \apj, 639, 1120

\bibitem[{{Kiss} {et~al.}(2011){Kiss}, {Mo{\'o}r}, {Szalai}, {Kov{\'a}cs},
  {Bayliss}, {Gilmore}, {Bienaym{\'e}}, {Binney}, {Bland-Hawthorn}, {Campbell},
  {Freeman}, {Fulbright}, {Gibson}, {Grebel}, {Helmi}, {Munari}, {Navarro},
  {Parker}, {Reid}, {Seabroke}, {Siebert}, {Siviero}, {Steinmetz}, {Watson},
  {Williams}, {Wyse}, \& {Zwitter}}]{Kiss:2011}
{Kiss}, L.~L., {et~al.} 2011, \mnras, 411, 117

\bibitem[{{Konopacky} {et~al.}(2010){Konopacky}, {Ghez}, {Barman}, {Rice},
  {Bailey}, {White}, {McLean}, \& {Duch{\^e}ne}}]{Konopacky:2010}
{Konopacky}, Q.~M., {Ghez}, A.~M., {Barman}, T.~S., {Rice}, E.~L., {Bailey},
  III, J.~I., {White}, R.~J., {McLean}, I.~S., \& {Duch{\^e}ne}, G. 2010, \apj,
  711, 1087

\bibitem[{{Kraus} {et~al.}(2014){Kraus}, {Shkolnik}, {Allers}, \&
  {Liu}}]{Kraus:2014}
{Kraus}, A.~L., {Shkolnik}, E.~L., {Allers}, K.~N., \& {Liu}, M.~C. 2014, \aj,
  147, 146

\bibitem[{{Lagrange} {et~al.}(2010){Lagrange}, {Bonnefoy}, {Chauvin}, {Apai},
  {Ehrenreich}, {Boccaletti}, {Gratadour}, {Rouan}, {Mouillet}, {Lacour}, \&
  {Kasper}}]{Lagrange2010}
{Lagrange}, A.-M., {et~al.} 2010, Science, 329, 57

\bibitem[{{Liu} {et~al.}(2012){Liu}, {Leggett}, {Golimowski}, {Chiu}, {Fan}, {Geballe}, {Schneider}, \& {Brinkmann}}]{Liu2012}
{Liu}, M.-C., {et~al.} 2012, \apj, 647, 1393

\bibitem[{{Manjavacas} {et~al.}(2014){Manjavacas}, {Bonnefoy}, {Schlieder}, {Allard}, {Rojo}, {Goldman}, {Chauvin}, {Homeier}, {Lodieu}, \& {Henning}}]{Manjavacas:2014}
{Manjavacas}, E., {et~al.} 2014, \aap, 564, 55

\bibitem[{{Malo} {et~al.}(2014{\natexlab{a}}){Malo}, {Artigau}, {Doyon},
  {Lafreni{\`e}re}, {Albert}, \& {Gagn{\'e}}}]{Malo:2014a}
{Malo}, L., {Artigau}, {\'E}., {Doyon}, R., {Lafreni{\`e}re}, D., {Albert}, L.,
  \& {Gagn{\'e}}, J. 2014{\natexlab{a}}, \apj, 788, 81

\bibitem[{{Malo} {et~al.}(2014{\natexlab{b}}){Malo}, {Doyon}, {Feiden},
  {Albert}, {Lafreni{\`e}re}, {Artigau}, {Gagn{\'e}}, \& {Riedel}}]{Malo:2014}
{Malo}, L., {Doyon}, R., {Feiden}, G.~A., {Albert}, L., {Lafreni{\`e}re}, D.,
  {Artigau}, {\'E}., {Gagn{\'e}}, J., \& {Riedel}, A. 2014{\natexlab{b}}, \apj,
  792, 37

\bibitem[{{Malo} {et~al.}(2013){Malo}, {Doyon}, {Lafreni{\`e}re}, {Artigau},
  {Gagn{\'e}}, {Baron}, \& {Riedel}}]{Malo:2013}
{Malo}, L., {Doyon}, R., {Lafreni{\`e}re}, D., {Artigau}, {\'E}., {Gagn{\'e}},
  J., {Baron}, F., \& {Riedel}, A. 2013, \apj, 762, 88

\bibitem[{{Marois} {et~al.}(2008){Marois}, {Macintosh}, {Barman}, {Zuckerman},
  {Song}, {Patience}, {Lafreni{\`e}re}, \& {Doyon}}]{Marois:2008}
{Marois}, C., {Macintosh}, B., {Barman}, T., {Zuckerman}, B., {Song}, I.,
  {Patience}, J., {Lafreni{\`e}re}, D., \& {Doyon}, R. 2008, Science, 322, 1348

\bibitem[{{Marois} {et~al.}(2010){Marois}, {Zuckerman}, {Konopacky},
  {Macintosh}, \& {Barman}}]{Marois:2010}
{Marois}, C., {Zuckerman}, B., {Konopacky}, Q.~M., {Macintosh}, B., \&
  {Barman}, T. 2010, \nat, 468, 1080

\bibitem[{{Marsh} {et~al.}(2013){Marsh}, {Wright}, {Kirkpatrick}, {Gelino},
  {Cushing}, {Griffith}, {Skrutskie}, \& {Eisenhardt}}]{Marsh:2013}
{Marsh}, K.~A., {Wright}, E.~L., {Kirkpatrick}, J.~D., {Gelino}, C.~R.,
  {Cushing}, M.~C., {Griffith}, R.~L., {Skrutskie}, M.~F., \& {Eisenhardt},
  P.~R. 2013, \apj, 762, 119

\bibitem[{{McLean} {et~al.}(2003){McLean}, {McGovern}, {Burgasser},
  {Kirkpatrick}, {Prato}, \& {Kim}}]{McLean:2003}
{McLean}, I.~S., {McGovern}, M.~R., {Burgasser}, A.~J., {Kirkpatrick}, J.~D.,
  {Prato}, L., \& {Kim}, S.~S. 2003, \apj, 596, 561

\bibitem[{{Mo{\'o}r} {et~al.}(2013){Mo{\'o}r}, {Szab{\'o}}, {Kiss}, {Kiss},
  {{\'A}brah{\'a}m}, {Szul{\'a}gyi}, {K{\'o}sp{\'a}l}, \& {Szalai}}]{Moor:2013}
{Mo{\'o}r}, A., {Szab{\'o}}, G.~M., {Kiss}, L.~L., {Kiss}, C.,
  {{\'A}brah{\'a}m}, P., {Szul{\'a}gyi}, J., {K{\'o}sp{\'a}l}, {\'A}., \&
  {Szalai}, T. 2013, \mnras, 435, 1376

\bibitem[{{Naud} {et~al.}(2014){Naud}, {Artigau}, {Malo}, {Albert}, {Doyon},
  {Lafreni{\`e}re}, {Gagn{\'e}}, {Saumon}, {Morley}, {Allard}, {Homeier},
  {Beichman}, {Gelino}, \& {Boucher}}]{Naud:2014}
{Naud}, M.-E., {et~al.} 2014, \apj, 787, 5

\bibitem[{{Padoan} \& {Nordlund}(2002)}]{Padoan:2002}
{Padoan}, P., \& {Nordlund}, {\AA}. 2002, \apj, 576, 870

%%%%%%%%%%%%%%%%%%%%%%%%%%%%%%%%%%
 
\bibitem[{{Pollack} {et~al.}(1996){Pollack}, {Hubickyj}, {Bodenheimer}, {Lissauer}, {Podolak}, M. \& {Greenzweig},}]{Pollack1996}
{Pollack}, J.~B., {Hubickyj}, O., {Bodenheimer}, P., {Lissauer}, J.~J., {Podolak}, M. \& {Greenzweig}, Y.
 1996, \icarus, 124, 62

%%%%%%%%%%%%%%%%%%%%%%%%%%%%%%%%%%

\bibitem[{{Reid} \& {Walkowicz}(2006)}]{Reid2006}
{Reid}, I.~N., \& {Walkowicz}, L.~M. 2006, \pasp, 118, 671



\bibitem[{{Rice} {et~al.}(2010){Rice}, {Barman}, {Mclean}, {Prato}, \&
  {Kirkpatrick}}]{Rice:2010}
{Rice}, E.~L., {Barman}, T., {Mclean}, I.~S., {Prato}, L., \& {Kirkpatrick},
  J.~D. 2010, \apjs, 186, 63

\bibitem[{{Rodriguez} {et~al.}(2013){Rodriguez}, {Zuckerman}, {Kastner},
  {Bessell}, {Faherty}, \& {Murphy}}]{Rodriguez:2013}
{Rodriguez}, D.~R., {Zuckerman}, B., {Kastner}, J.~H., {Bessell}, M.~S.,
  {Faherty}, J.~K., \& {Murphy}, S.~J. 2013, \apj, 774, 101

\bibitem[{{Saumon} \& {Marley}(2008)}]{Saumon:2008}
{Saumon}, D., \& {Marley}, M.~S. 2008, \apj, 689, 1327

\bibitem[{{Simcoe} {et~al.}(2013){Simcoe}, {Burgasser}, {Schechter}, {Fishner},
  {Bernstein}, {Bigelow}, {Pipher}, {Forrest}, {McMurtry}, {Smith}, \&
  {Bochanski}}]{Simcoe:2013}
{Simcoe}, R.~A., {et~al.} 2013, \pasp, 125, 270

\bibitem[{{Skrutskie} {et~al.}(2006){Skrutskie}, {Cutri}, {Stiening},
  {Weinberg}, {Schneider}, {Carpenter}, {Beichman}, {Capps}, {Chester},
  {Elias}, {Huchra}, {Liebert}, {Lonsdale}, {Monet}, {Price}, {Seitzer},
  {Jarrett}, {Kirkpatrick}, {Gizis}, {Howard}, {Evans}, {Fowler}, {Fullmer},
  {Hurt}, {Light}, {Kopan}, {Marsh}, {McCallon}, {Tam}, {Van Dyk}, \&
  {Wheelock}}]{Skrutskie:2006}
{Skrutskie}, M.~F., {et~al.} 2006, \aj, 131, 1163

\bibitem[{{Song} {et~al.}(2002){Song}, {Bessell}, \& {Zuckerman}}]{Song:2002}
{Song}, I., {Bessell}, M.~S., \& {Zuckerman}, B. 2002, \apjl, 581, L43

\bibitem[{{Spiegel} {et~al.}(2011){Spiegel}, {Burrows}, \&
  {Milsom}}]{Spiegel:2011}
{Spiegel}, D.~S., {Burrows}, A., \& {Milsom}, J.~A. 2011, \apj, 727, 57

\bibitem[{{Torres} {et~al.}(2000){Torres}, {da Silva}, {Quast}, {de la Reza},
  \& {Jilinski}}]{Torres:2000}
{Torres}, C.~A.~O., {da Silva}, L., {Quast}, G.~R., {de la Reza}, R., \&
  {Jilinski}, E. 2000, \aj, 120, 1410

\bibitem[{{Torres} {et~al.}(2008){Torres}, {Quast}, {Melo}, \&
  {Sterzik}}]{Torres:2008}
{Torres}, C.~A.~O., {Quast}, G.~R., {Melo}, C.~H.~F., \& {Sterzik}, M.~F. 2008,
  {Young Nearby Loose Associations}, ed. B.~{Reipurth}, 757

\bibitem[{{Veras} {et~al.}(2009){Veras}, {Crepp}, \& {Ford}}]{Veras:2009}
{Veras}, D., {Crepp}, J.~R., \& {Ford}, E.~B. 2009, \apj, 696, 1600

\bibitem[{{Vorobyov}(2013)}]{Vorobyov:2013}
{Vorobyov}, E.~I. 2013, \aap, 552, A129

\bibitem[{{Wright} {et~al.}(2010){Wright}, {Eisenhardt}, {Mainzer}, {Ressler},
  {Cutri}, {Jarrett}, {Kirkpatrick}, {Padgett}, {McMillan}, {Skrutskie},
  {Stanford}, {Cohen}, {Walker}, {Mather}, {Leisawitz}, {Gautier}, {McLean},
  {Benford}, {Lonsdale}, {Blain}, {Mendez}, {Irace}, {Duval}, {Liu}, {Royer},
  {Heinrichsen}, {Howard}, {Shannon}, {Kendall}, {Walsh}, {Larsen}, {Cardon},
  {Schick}, {Schwalm}, {Abid}, {Fabinsky}, {Naes}, \& {Tsai}}]{Wright:2010}
{Wright}, E.~L., {et~al.} 2010, \aj, 140, 1868

\bibitem[{{Yee} \& {Jensen}(2010)}]{Yee:2010}
{Yee}, J.~C., \& {Jensen}, E.~L.~N. 2010, \apj, 711, 303

\bibitem[{{Zacharias} {et~al.}(2005){Zacharias}, {Monet}, {Levine}, {Urban},
  {Gaume}, \& {Wycoff}}]{Zacharias:2005}
{Zacharias}, N., {Monet}, D.~G., {Levine}, S.~E., {Urban}, S.~E., {Gaume}, R.,
  \& {Wycoff}, G.~L. 2005, VizieR Online Data Catalog, 1297, 0

\bibitem[{{Zuckerman} \& {Song}(2004)}]{Zuckerman:2004}
{Zuckerman}, B., \& {Song}, I. 2004, \araa, 42, 685

\bibitem[{{Zuckerman} {et~al.}(2001){Zuckerman}, {Song}, \&
  {Webb}}]{Zuckerman:2001}
{Zuckerman}, B., {Song}, I., \& {Webb}, R.~A. 2001, \apj, 559, 388

\bibitem[{{Zuckerman} \& {Webb}(2000)}]{Zuckerman:2000}
{Zuckerman}, B., \& {Webb}, R.~A. 2000, \apj, 535, 959

\end{thebibliography}
\end{document}